\documentstyle[11pt,aaspp4,rotate]{article}

\def\lya{Ly$\alpha$ }
\def\lyb{Ly$\beta$ }
\def\lyc{Ly$\gamma$ }
\def\lyd{Ly$\delta$ }
\def\strom{Str\"{o}mgren}

\begin{document}

\slugcomment{AJ, in press}
\title{Constraining the Evolution of the Ionizing Background and the Epoch of Reionization with $z\sim 6$ Quasars II: A Sample of 19 Quasars\altaffilmark{1,2}}

\author{Xiaohui Fan\altaffilmark{\ref{Arizona}},
Michael A. Strauss\altaffilmark{\ref{Princeton}},
Robert H. Becker\altaffilmark{\ref{UCDavis},\ref{IGPP}},
Richard L. White\altaffilmark{\ref{STScI}}, 
James E. Gunn\altaffilmark{\ref{Princeton}},
Gillian R. Knapp\altaffilmark{\ref{Princeton}},
Gordon T. Richards\altaffilmark{\ref{Princeton},\ref{JHU}},
Donald P. Schneider\altaffilmark{\ref{PSU}},
J. Brinkmann\altaffilmark{\ref{APO}}\\
and Masataka Fukugita\altaffilmark{\ref{CosmicRay}}
}

\altaffiltext{1}{Based on observations obtained with the
Sloan Digital Sky Survey;
at the W.M. Keck Observatory, which is operated as a scientific partnership
among the California Institute of Technology, the University of California and the National Aeronautics and
Space Administration, made possible by the generous financial support of the W.M. Keck 
Foundation, with the MMT Observatory, a joint facility of the University of
Arizona and the Smithsonian Institution,
and with the Kitt Peak National Observatory 4-meter Mayall Telescope.}
\altaffiltext{2}{This paper is dedicated to the memory of John N. Bahcall.}

\newcounter{address}
\setcounter{address}{3}
\altaffiltext{\theaddress}{Steward Observatory, The University of Arizona,
Tucson, AZ 85721; fan@as.arizona.edu
\label{Arizona}}
\addtocounter{address}{1}
\altaffiltext{\theaddress}{Princeton University Observatory, Princeton,
NJ 08544
\label{Princeton}}
\addtocounter{address}{1}
\altaffiltext{\theaddress}{Physics Department, University of California, Davis,
CA 95616
\label{UCDavis}}
\addtocounter{address}{1}
\altaffiltext{\theaddress}{IGPP/Lawrence Livermore National Laboratory, Livermore,
CA 94550
\label{IGPP}}
\addtocounter{address}{1}
\altaffiltext{\theaddress}{Space Telescope Science Institute, Baltimore, MD 21218
\label{STScI}}
\addtocounter{address}{1}
\altaffiltext{\theaddress}{
Department of Physics and Astronomy, The Johns Hopkins University,
Baltimore, MD 21218
\label{JHU}}
\addtocounter{address}{1}
\altaffiltext{\theaddress}{Department of Astronomy and Astrophysics,
The Pennsylvania State University,
University Park, PA 16802
\label{PSU}}
\addtocounter{address}{1}
\altaffiltext{\theaddress}{Apache Point Observatory, P. O. Box 59,
Sunspot, NM 88349-0059
\label{APO}}
\addtocounter{address}{1}
\altaffiltext{\theaddress}{Institute for Cosmic Ray Research, University of
Tokyo, Midori, Tanashi, Tokyo 188-8502, Japan
\label{CosmicRay}}

\begin{abstract}
We study the evolution of the ionization state of the intergalactic medium~(IGM)
at the end of the reionization epoch using moderate resolution
spectra of a sample of nineteen
quasars at $5.74 < z_{\rm em} < 6.42$ discovered in the Sloan Digital Sky Survey.
Three methods are used to trace IGM properties:
(a) the evolution of the Gunn-Peterson (GP) optical depth in the Ly$\alpha$, $\beta$, and $\gamma$ transitions;
(b) the distribution of lengths of dark absorption gaps, and (c) the size of HII regions around
luminous quasars.
Using this large sample, we find that the evolution of the ionization state of the IGM accelerated at $z>5.7$:
the GP optical depth evolution changes from $\tau^{\rm eff}_{\rm GP} \sim (1+z)^{4.3}$
to $(1+z)^{\gtrsim 11}$, and the average length of dark gaps with $\tau>3.5$
increases from $<10$ to $>80$ comoving Mpc.
The dispersion of IGM properties along different lines of sight
also increases rapidly, 
implying 
fluctuations by a factor of $\gtrsim 4$ in the UV background at $z>6$, when
the mean free path of UV photons is comparable
to the correlation length of the star forming galaxies that are thought to have caused reionization.
The mean length of dark gaps
shows the most dramatic increase at $z\sim 6$, as well as the largest
line-of-sight variations.
We suggest using dark gap statistics as  a powerful probe of
the ionization state of the IGM at yet higher redshift.
The sizes of HII regions around luminous quasars decrease rapidly
towards higher redshift, 
suggesting that the neutral fraction
of the IGM has increased by a factor of $\gtrsim 10$ from $z=5.7$ to 6.4,
consistent with the value derived from the GP optical depth.
The mass-averaged neutral fraction is $1 - 4$\% at $z\sim 6.2$
based on the GP optical depth and HII region size measurements.
The observations suggest that $z\sim 6$ is the end of the overlapping
stage of reionization, and 
are inconsistent with a mostly neutral
IGM at $z\sim 6$, as indicated by the finite length of dark absorption gaps.
\end{abstract}


\section{Introduction}

After the recombination epoch at $z\sim 1100$, the universe became 
mostly neutral, until the first generation of stars and quasars
reionized the intergalactic medium (IGM) and ended the
cosmic ``dark ages'' (e.g., Rees 1998).
Cosmological models predict reionization 
at redshifts between 6 and 20 (e.g., Gnedin \& Ostriker 1997,
Gnedin 2000, 2004, Ciardi et al. 2001, Benson et al. 2001, Razoumov et al. 2002, Cen 2003a,b, Ciardi, Ferrara \& White 2003, Haiman \& Holder 2003, Wyithe \& Loeb 2003a, Somerville et al. 2003,
\cite{choudhury05}). 
When and how the universe reionized remains one of the fundamental
questions of modern cosmology (for reviews of theoretical models, see
Barkana \& Loeb 2001, Loeb \& Barkana 2001, and
Ciardi \& Ferrara 2005).

The last few years have witnessed the first direct observational constraints
on the history of reionization.
Lack of complete Gunn-Peterson (GP, 1965) absorption at $z<6$ indicates
that the IGM is highly ionized by that epoch
(e.g. Fan et al. 2000, 2001, Becker et al. 2001, Djorgovski et al. 2001, Songaila \&
Cowie 2002). 
Spectroscopic observations of the highest-redshift quasars and galaxies
provide a number of observational constraints suggesting that $z\sim 6$
indeed marks the end of the reionization epoch:
\begin{itemize}
\item The detection of complete GP absorption troughs in
the spectra of quasars at $z>6$ suggests that
the neutral fraction of the IGM is increasing rapidly with redshift (Cen \& McDonald 2002,
Fan et al. 2002, \cite{lidz2002}, White et al. 2003, \cite{gnedin2004}).
This feature is consistent with the phase
of IGM evolution at which  individual HII regions overlapped;
at this point,  the IGM ionization state evolved rapidly, similar to a phase transition
(cf. Songaila 2004 and discussion below). The lower limit of the 
volume averaged neutral fraction is $\sim 10^{-3}$ based on Gunn-Peterson
measurements, although it could be much higher in the deepest troughs.
\item The sizes of large, highly ionized regions around luminous quasars at 
$z>6$
are consistent with those predicted for a quasar ``\strom\ sphere'' expanding
into a surrounding largely neutral IGM (\cite{mesinger04a}, \cite{mesinger04b},
\cite{wyithe04a}). 
The neutral fraction of the IGM is estimated to be as high as $>20$\% along
some lines of sight, considerably higher than the lower limit using the Gunn-Peterson
optical depth. But these measurements are made uncertain because of
our lack of knowledge 
of quasar lifetimes, bolometric corrections of quasar luminosities, the bias factor
and the clumpiness of the IGM around high redshift quasars (e.g., \cite{yu05a}, \cite{yu05b}).
\item The observed temperature evolution of the IGM at $z\sim 3-5$ suggests that the reionization
epoch is not much higher than 8 (\cite{hui03}). In models with
reionization at much higher redshift, the IGM cools to a much lower temperature
than that observed at $z\sim 2 - 4$.
\end{itemize}
Quasars and AGNs are unlikely to provide enough UV photons to reionize the
universe at $z>6$ (e.g. Fan et al. 2001, 
\cite{dijkstra04}, \cite{meiksin05}).
Star forming galaxies at early epochs are the most likely candidates to
ionize the IGM, although current observations still have poor constraints
on the total ionizing photon budget at $z\sim 6$ (\cite{yan04b}, \cite{stiavelli04}).
Recent measurements of the luminosity density of
Lyman break galaxies using deep HST observations
show a moderate decline in total UV luminosity of star forming galaxies 
at $z>6$ (\cite{bunker04}, \cite{yan04b}, \cite{bouwens04a},b). 
If the reionization process was similar to a phase transition, 
the high-redshift quasar observations suggest that the Universe could have been mostly neutral
as late as $z = 6 - 8$.

Polarization measurements of the cosmic microwave background
(CMB) based on WMAP first-year data (Kogut et al. 2003) show a large
optical depth due to Thompson scattering of electrons ($\tau = 0.17 \pm 0.04$) 
in the early 
Universe, suggesting that the IGM was largely ionized
by $z\sim 17\pm 4$.
WMAP three-year data (Spergel et al. 2006) show a smaller optical depth
($\tau = 0.09 \pm 0.03$) based on the new EE measurements, indicating a larely ionized IGM by 
$z\sim 10 \pm 3$.
The WMAP polarization results, combined with low redshift constraints,
suggest that reionization could be an extended process rather than
a phase transition (Cen 2003a, b, \cite{wyithe03a},
\cite{haiman03}, \cite{somerville03b}, \cite{gnedin2004}, \cite{choudhury05}).
Even at $z\sim 6$, several lines of evidence suggest that a uniform end of reionization
with  a clean phase transition in the ionization state is too simplistic:
\begin{itemize}
\item While the optical depth no doubt has increased at $z>6$, it is not agreed
whether the increase is similar to a sharp phase transition (e.g., Fan et al. 2002),
or is simply a continuation of the gradual thickening of the Ly$\alpha$ forest
(e.g. Songaila 2004).
Although optical depth measurements of quasar spectra are consistent among
different studies, the different 
interpretations arise because: (1) complete Gunn-Peterson troughs
give only a lower limit to the  optical depth; (2) there are complications
in the interpretation of  the Ly$\beta$ measurements; and (3) there is a lack of a clear definition
of ``gradual'' vs. ``sharp'' transitions.
\item There are large line of sight variations from one quasar to another. The GP trough detected 
in the spectrum of SDSS J1030+0524 ($z=6.28$, Fan et al. 2001, Becker et al. 2001,
White et al. 2003) shows complete absorption in the Ly$\alpha$, $\beta$ and $\gamma$
transitions. However,
the line of sight of SDSS J1148+5251 ($z=6.42$, Fan et al. 2002)
shows clear transmission, especially in the Ly$\beta$ and $\gamma$ transitions
(White et al. 2003, 2005, \cite{OF2005}), indicating that the IGM along
this line of sight is still more than 99\% ionized.
\item The lack of evolution in the luminosity functions of Ly$\alpha$
emitting galaxies from $z\sim 5.7$ to $z\sim 6.5$ is also consistent with
the picture that the IGM is largely ionized at $z\sim 6$; in a largely
neutral IGM, an extended Gunn-Peterson absorption damping wing would attenuate 
the Ly$\alpha$ emission of high-redshift galaxies
(\cite{hu04}, 2005, \cite{malhotra04}, \cite{stern05}, \cite{malhotra06}). 
However, due to uncertainties in the clustering of Ly$\alpha$ galaxies
and the clumpiness of the IGM, a relatively high neutral fraction cannot yet be ruled out
by the \lya emitter observations
(\cite{santos04}, \cite{haiman05}, \cite{wyithe05}, \cite{furlanetto05}).
\end{itemize}

Reionization is a complex process which results from the interplay of
structure formation, early star formation and feedback, and radiative transfer
in a clumpy IGM. Recent developments in both semi-analytic models and 
simulations aim at understanding these processes in the context of 
improved observational constraints and future observations.
Models that include feedback from the formation of the earliest, metal-free populations
can generally fit both the Gunn-Peterson measurements at low redshift and
a high CMB polarization measurement 
(e.g., Cen 2003a, b, Wyithe \& Loeb 2003a, Haiman \& Holder 2003, \cite{melchiorri05}).
Cosmological simulations with improved resolution, box size and treatment
of radiative transfer (e.g. Gnedin 2000, 2004, Razoumov et al.\ 2002, \cite{pn05}) 
are beginning to make realistic comparison with
observations possible. 

To make further progress observationally, spectroscopy of a large sample
of luminous sources is needed to explore the line of sight variation in the
GP optical depth measurements expected due to clumpiness in the IGM and
clustering of ionizing sources (e.g. Wyithe \& Loeb 2004b, \cite{pn05}).
Fan et al. (2002, Paper I) presented measurements of the ionization state 
of the IGM at $z\sim 6$ and constraints on the epoch of reionization, using
a sample of four quasars at $z=5.74 - 6.28$.
In this follow-up paper, we describe measurements of Lyman absorption using spectroscopic
observations of a  sample of nineteen quasars
at $z=5.74 - 6.42$, discovered from the imaging data
of the Sloan Digital Sky Survey (Fan et al. 2000, 2001, 2002, 2003, 2004, 2006).
We address the following three questions:
\begin{enumerate}
\item  How fast do  {\bf the average and the dispersion} of the IGM ionization state
 evolve in the redshift range $z = 5 - 6.4$? 
In particular, is there a well-defined break in these quantities as 
a function of redshift?

\item Is the IGM consistent with
a large neutral fraction at least along some lines of sight? 
What is the {\em upper} limit of the neutral fraction when considering
all the observational constraints?

\item Can we define a set of statistics from the observed spectrum
that are also easily extractable from cosmological simulations and 
allow direct comparison with theoretical models (e.g., Songaila \& Cowie 2002,
\cite{pn05})?
\end{enumerate}

The paper is organized as follows:
in \S2, we describe our  spectroscopic sample of 19 SDSS quasars at $z=5.74 - 6.42$.
In \S3, we measure the evolution of Gunn-Peterson
optical depths. We present \lya measurements in \S3.1, discuss in detail the optimal way of combining
Ly$\alpha$, $\beta$ and $\gamma$ measurements in \S3.2, and present
the measurement of the dispersion of optical depth along different lines
of sight in \S3.3.
In \S4, we derive a number of quantities characterizing the ionization
state of the IGM: the evolution of the ionizing background based on
a photoionization model (\S4.1), and of the mean free path of
ionizing photons and the neutral fraction of the IGM (\S4.2), following
the methods developed in Paper I.  We compare these results with simulations in \S4.3.
In \S5, we study the 
presence of complete Gunn-Peterson absorption troughs in quasar spectra (\S5.1),
and discuss in detail the statistics of the dark gap distribution (\S 5.2). 
In \S5.3, we use these statistics to place an independent upper limit
on the IGM neutral fraction.
In \S6, we measure the sizes of HII regions around these 19 quasars,
and use their evolution to place an independent
constraint on the rate of neutral fraction evolution at $z\sim 6$.
Finally, we summarize the results in \S7.
Through the paper, we use the WMAP cosmology
($\rm H_0 = 71\ km\ s^{-1}\ Mpc^{-1}$, $\Omega_{\Lambda} = 0.73$,
$\Omega_{M} = 0.27$, and $\Omega_b = 0.04$, Spergel et al. 2003) to present our results.

\begin{figure}
\epsscale{0.9}
\plotone{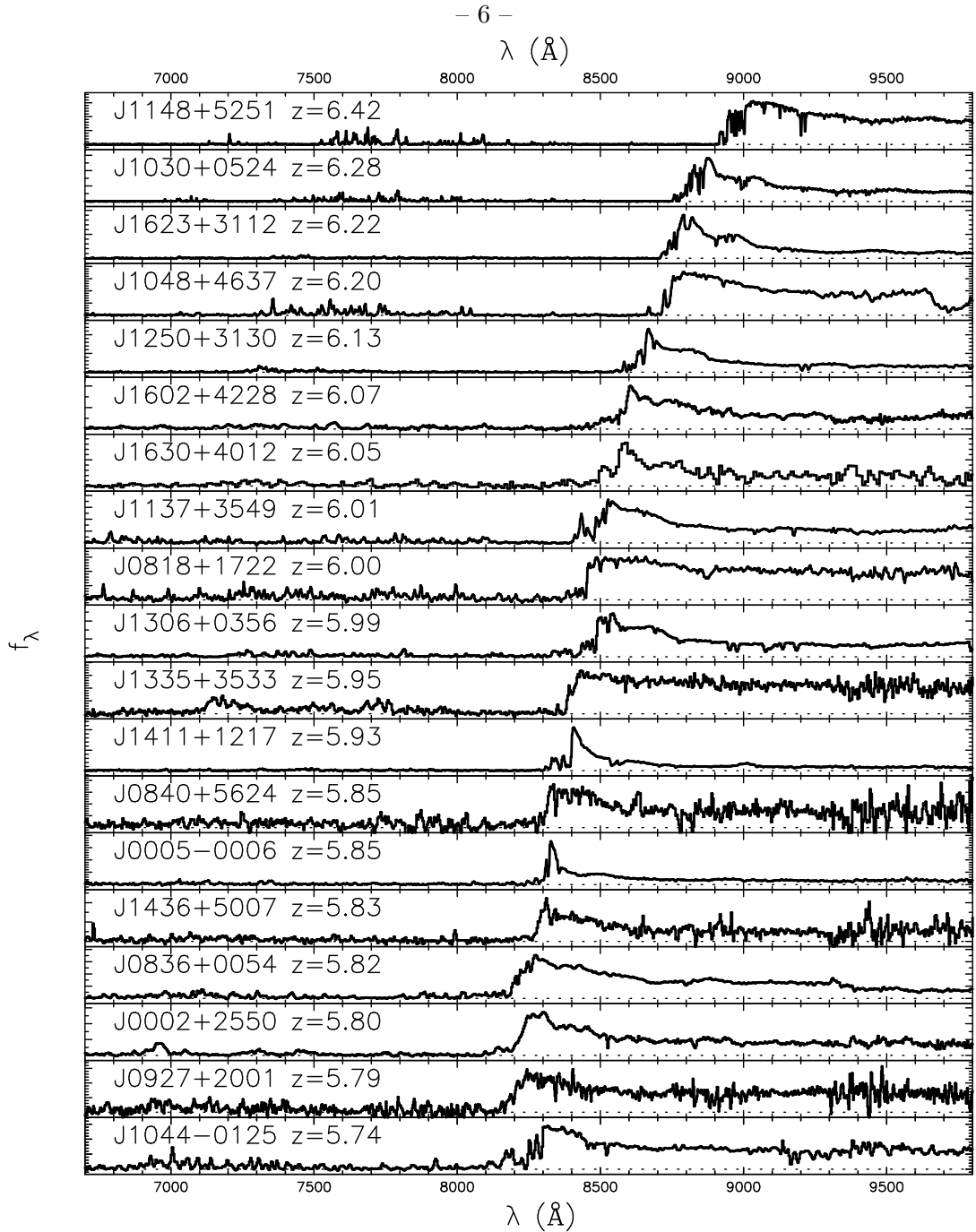}
\caption{Spectra of our sample of nineteen SDSS quasars at $5.74 < z < 6.42$.
Twelve of the spectra were taken with Keck/ESI, while the others 
were observed with
the MMT/Red Channel and Kitt Peak 4-meter/MARS spectrographs. See Table 1 for
detailed information.} 
\end{figure}

\section{A Spectroscopic Sample of $z\sim 6$ Quasars}

The Sloan Digital Sky Survey (York et al. 2000) is using
a dedicated 2.5m telescope (Gunn et al. 2006) and a large format CCD camera (\cite{Gunnetal})
at the Apache Point Observatory in New Mexico
to obtain images in five broad bands ($u$, $g$, $r$, $i$ and $z$,
centered at 3551, 4686, 6166, 7480 and 8932 \AA, respectively; \cite{F96})
of high Galactic latitude sky in the Northern Galactic Cap.
About $7500$ deg$^2$ of sky in the Northern Galactic Cap have been imaged at the
time of this writing (Oct 2005).
The imaging data are processed by the astrometric pipeline
(\cite{Astrom}) and the photometric pipeline (\cite{Photo}),
and are photometrically calibrated to a standard star network
(\cite{Smith02}; see also \cite{Hogg01}, Ivezic et al. 2004, Tucker et al. 2005).

Quasars with redshift greater than 5.7 are characterized by very red $i-z$ colors.
Since 2000, we have been using the imaging data from the SDSS
to select $i$-dropout candidates.
At $z>5.7$, all quasars have deep Ly$\alpha$ absorption
and satisfy our color-selection criteria
($i-z >2.2$); the color-selection will
not bias IGM measurements.
Because such quasars are very red and faint, and can be confused with 
cosmic rays and brown dwarfs, 
we do not use the automated
SDSS quasar selection algorithm (Richards et al. 2002a) to select and target them.
In a series of papers (Fan et al. 2000, 2001, 2003, 2004, 2006),
we have reported the selection and identification of nineteen quasars
selected from about 6600 deg$^2$ of SDSS imaging.
The sample covers the redshift range  $z=5.74 - 6.42$,
and a range of absolute magnitudes in rest-frame 1450\AA,
$M_{1450} = -26.2$ to $-27.9$. They are the most distant
luminous quasars published to date (see also Mahabal et al. 2005, R. Cool et al. in preparation).

Table 1 summarizes the properties and spectroscopic observations of these quasars.
These spectra were taken with a number of different spectrographs and
exposure times, resulting in a range of signal-to-noise ratio (S/N)
 and spectral resolution. 
Twelve objects in the sample have spectra taken with
the Echellette Imaging Spectrograph
(ESI, \cite{ESI}) on the Keck II telescope, of which four are presented in this paper for the
first time
(see White et al. 2003 for details of observations and data reduction). 
These data have spectral resolution $R\sim 3000 - 6000$,
depending on the seeing and slit used.
To present uniform data, we have binned all the Keck/ESI spectra
to a resolution of $R = 2600$.
Although the S/N of these spectra vary by a factor of $\sim 7$,
we have sufficiently long observations for all quasars at $z>6.1$ to 
ensure proper measurement of complete Gunn-Peterson troughs. 
For quasars at $z<6.1$, and absorption redshift $z_{abs} < 5.7$,
the IGM transmission is well detected even with relatively short exposures.
In most of this paper, we will use averaged transmission and effective
optical depth as the main observables.
The non-uniform S/N among the lower redshift quasars is not a limiting
factor in the analysis: 
in fact, uncertainties in the average transmission are dominated
by large sample variance (\S3.3).
Thus our data are adequate for our purposes, although calculating higher-order
statistics requires more uniformly high S/N spectra,
such as presented by White et al. (2003), Songaila (2004) and
Becker et al. (2006).

\section{Evolution of Gunn-Peterson Optical Depth}

Figure 1 presents the spectra of the 19 quasars.
At least two of the objects (SDSS J1048+4637
\footnote{Throughout this paper, we use the shortened version
of quasar name, SDSS Jhhmm+ddmm. Complete IAU names are given in
Table 1.}, $z=6.20$,
Maiolino et al. 2004, and SDSS J1044-0125, $z=5.74$, Djorgovski et al. 2001,
Goodrich et al. 2001)
are broad absorption line (BAL) quasars. In analyzing their spectra,
we have excluded regions that are affected by the Ly$\alpha$ and Ly$\beta$ BAL features.
Our optical spectroscopy is usually adequate to detect SiIV BAL troughs.
However, at $z\sim 6$, the strongest CIV BAL troughs are redshifted
to $\lambda \gtrsim 1 \mu$m. Fewer than half of the objects in the sample
have adequate IR spectroscopic observations to detect the CIV BAL feature. 
Note that the fraction of BAL quasars at $z<4$ is $\sim 15$\% in color-selected
samples (e.g. \cite{reinhart2003}).

Figure 1 shows strong redshift evolution of the transmission of the IGM:
transmitted flux is clearly detected in the spectra of quasars at $z<6$
and blueward of the \lya emission line; the absorption troughs deepen for the high-redshift
quasars, and complete Gunn-Peterson absorption begins to appear along 
lines of sight at $z>6.1$. In this section, we present detailed measurements
of the average Gunn-Peterson optical depth in the Ly$\alpha$, $\beta$ and
$\gamma$ transitions
and discuss the variation of the optical depth along different lines of sight. 

\subsection{Ly$\alpha$}

The Gunn-Peterson (1965) optical depth to \lya photons is
\begin{equation}
\tau_{\rm GP} = \frac{\pi e^2}{m_e c} f_{\alpha} \lambda_{\alpha} H^{-1}(z) n_{\rm HI },
\end{equation}
where $f_{\alpha}$ is the oscillator strength of the \lya transition,
$\lambda_\alpha$ = 1216\AA, $H(z)$ is the Hubble constant at redshift
$z$, and $n_{\rm HI }$ is the density of neutral hydrogen in the IGM.
At high redshifts, using $H(z) \propto  h\Omega_m^{1/2}(1+z)^{3/2}$, and 
converting hydrogen density to the present-day baryon density parameter
$\Omega_b$, the GP
optical depth for a uniformly distributed IGM can be re-written as:
\begin{equation}
\tau_{\rm GP} (z) = 1.8 \times 10^5 h^{-1} \Omega_m^{-1/2}
\left( \frac{\Omega_b h^2}{0.02} \right)
\left ( \frac{1+z}{7} \right )^{3/2}
\left( \frac{n_{\rm HI}}{\langle n_{\rm H}}\rangle \right ).
\end{equation}
We define the transmitted flux ratio as the average
ratio of observed flux to that in the absence of absorption:
\begin{equation}
{\cal T}(z_{\rm abs}) \equiv \left\langle f_\nu^{\rm obs}/f_\nu^{\rm int} \right\rangle,
\end{equation}
where the average is over a certain wavelength or redshift range along
the line of sight.
We also define the effective GP optical depth,  
\begin{equation}
\tau_{\rm GP}^{\rm eff}  \equiv - \ln({\cal T}).
\end{equation}
Note that for a clumpy IGM, 
most of the flux is transmitted through regions with low $n_{\rm HI}$,
while high density regions simply saturate, resulting in a much smaller
effective optical depth than the average of the
pixel-to-pixel optical depth,  $\tau^{\rm eff} < \langle \tau \rangle$
(for further discussion, see Paper I, Oh \& Furlanetto 2005,
and \S4 of this paper).

We measure the transmitted flux ratio in the \lya transition
as a function of redshift for the sample
of 19 quasars as follows:
(1) we define the maximum redshift $z_{\rm max}^{\alpha}$ as the 
maximum \lya absorption redshift which is not affected the proximity
effect from the quasar itself (\S6). 
For the two known BAL quasars, regions affected by the BAL are excluded
based on their SiIV and CIV absorption redshifts. For non-BAL quasars,
we find $z_{\rm em}-z_{\rm max}^{\alpha} = 0.07 - 0.23$, with the difference
increasing towards lower redshift and higher luminosity (\S6).
(2) The minimum absorption redshift considered is chosen as
$1+z_{\rm min}^{\alpha} = (1+z_{\rm em})\times1040 \rm \AA/\lambda_{\alpha}$, 
where 1040 \AA\ is the shortest wavelength
that is {\em not} affected by the Ly$\beta$+OVI emission line.
The redshift interval covered by each line of sight ranges from 0.76 to 0.98.
(3) We choose a redshift interval $\Delta z  = 0.15$ for the calculation of
the average in Eq.~(3).
This interval corresponds to $\sim 60$ Mpc in co-moving distance at $z\sim 6$,
larger than any large-scale structure at that redshift
(see \S4.1). It is also sufficiently 
wide that the measurement error due to photon noise is always small in the absence
of complete GP troughs.
(4) We assume an underlying 
intrinsic spectrum with a power law continuum
$f_{\nu} \propto \nu^{-0.5}$ (Vanden Berk et al. 2001)
and a double Gaussian \lya+NV emission line.
The transmitted flux is measured by extrapolating this intrinsic spectrum $f_\nu^{\rm int}$
to shorter wavelength.
When $\cal{T}$ approaches zero at high redshift, 
the optical depth measurement depends only logarithmically 
on the exact shape of the power law continuum.
As discussed in detail in Paper I, the dominant error of the optical depth
measurement is sample variance in different regions of the IGM.
The \lya+NV emission line is only important for the region closest to
the \lya emission line in the quasar proximity zone (\S6).
For the whole sample, the \lya transmitted flux measurements cover a 
redshift range of $4.9 < z_{\rm abs} < 6.3$, with a total length of 14.6 in 
redshift ($\sim 6000$ comoving Mpc) when adding up all lines of sight.

\begin{figure}
.
\epsscale{1.0}
\plotone{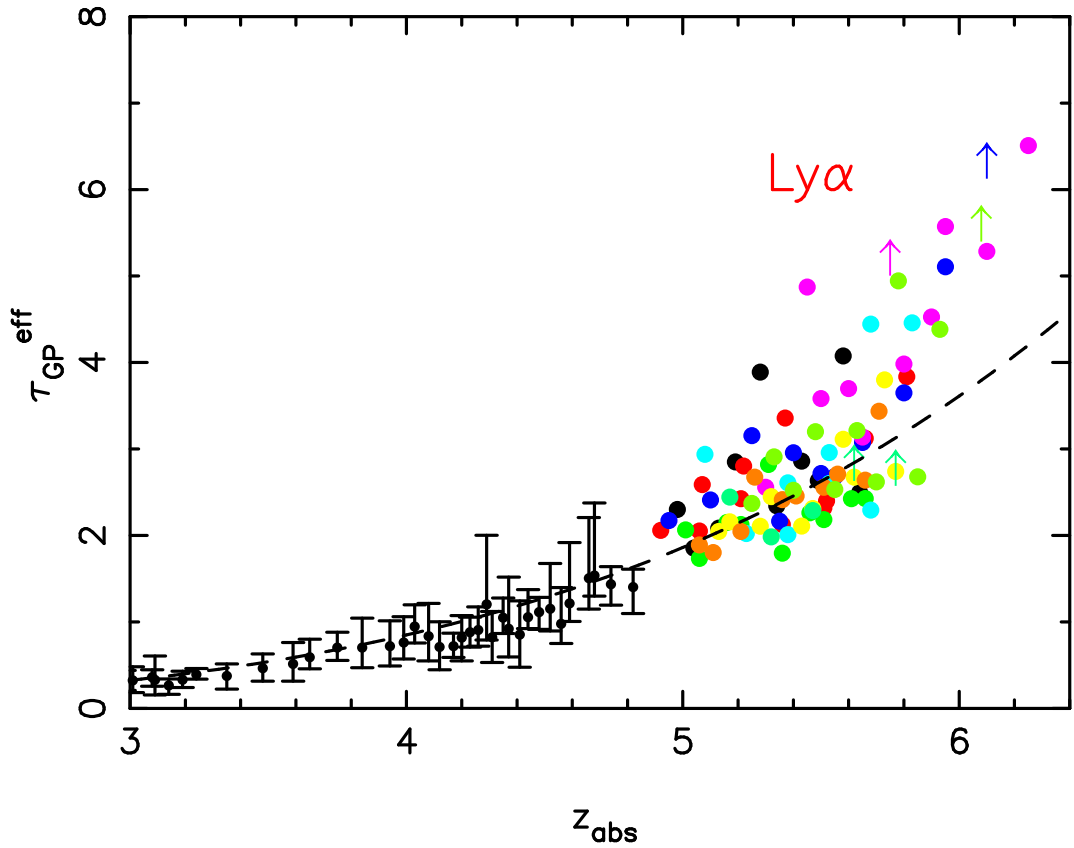}
\caption{Evolution of the Ly$\alpha$ Gunn-Peterson optical depth with redshift
averaged over intervals $\Delta z = 0.15$ along each line of sight.
At $z_{\rm abs} = 4.8 - 6.3$, the sample of 19 quasars in this paper yields
a total of 97 independent measurements covering a total redshift interval
of $\Delta z = 14.6$ (large symbols). 
In  complete GP troughs in which no flux is detected, the 2-$\sigma$ lower limit
on optical depth is indicated with an arrow.
We also include the measurements at lower redshift from
Songaila (2004, small symbols). The dashed curve shows the best-fit power-law for $z_{abs} < 5.5$:
$\tau_{\rm GP}^{\rm eff} = (0.85\pm0.06) \times \left(\frac{1+z}{5}\right)^{4.3\pm0.3}$. At $z_{\rm abs} \gtrsim 5.7$,
the evolution accelerates,  with increased dispersion and a rapid deviation from the power-law relation.}
\end{figure}

Table 2 presents the measurements of the \lya transmitted flux ratio.
Figure 2 presents the effective optical depth for the \lya transition as a
function of redshift.
When flux is detected at less than the 2-$\sigma$ level, a 2-$\sigma$ lower limit
on the optical depth is used.
At $z_{\rm abs} < 4.9$, we plot the data from Songaila (2004), where 95\% variation
due to sample variance is shown for each redshift.
The dashed curve shows the best-fit power-law for $z_{\rm abs} < 5.5$:
\begin{equation}
\tau_{\rm GP}^{\rm eff} = 0.85 \times \left(\frac{1+z}{5} \right)^{4.3}. 
\end{equation}
The results here are consistent with those in previous papers,
now using a significantly larger sample at $z_{\rm abs} > 5.7$. 
While at $z_{\rm abs} < 5.7$, the evolution of optical depth closely follows  the $(1+z)^{4.3}$
power law, there appears to be an accelerated evolution at
$z_{\rm abs} \gtrsim 5.7$, where the optical depth rapidly exceeds a simple
extrapolation from lower redshift. 

In addition to the increase in the average \lya optical depth, there is also 
an increase in the {\em dispersion} of this measurement. 
Dark regions with $\tau_{\rm GP}^{\rm eff} > 5.5$ occur as
early as $z_{\rm abs} \sim 5.7$. 
Deep, complete GP absorption troughs are detected in three different lines
of sight with various redshifts and lengths (\S4.2).
For example, at $z_{\rm abs} > 5.9$, we have eight independent measurements, resulting in
two upper limits ($\tau_{\rm GP}^{\rm eff} > 6-7$) and six detections
($\tau_{\rm GP}^{\rm eff} = 3.9 - 6.5$). The smallest optical depth measured
at $z>6$  does not
deviate significantly from the low-redshift extrapolation (see e.g. Songaila 2004).
We discuss the increasing dispersion of optical depth in \S3.3.

\subsection{Ly$\beta$ and Ly$\gamma$}

At a given absorption redshift, the GP optical depth $\tau_{\rm GP} \propto f\lambda$, where $f$ and $\lambda$
are the oscillator strength and rest-frame wavelength of the transition.
For the same neutral hydrogen density, the GP optical depths of \lyb and \lyc are
factors of 6.2 and 17.9 smaller than that of Ly$\alpha$, respectively. When \lya absorption
saturates at $z>6$, \lyb and \lyc
can in principle provide more stringent constraints on the IGM ionization state.
However, just as we have seen that the average and effective GP optical depths
are different, in a clumpy IGM, the ratio of {\em effective} optical depths
between \lya and \lyb/~\lyc transition are {\em smaller} than 
the factors above. The \lyb/\lyc absorptions in the quasar
spectra are further affected by the lower order transitions in the foreground
(at lower $z_{\rm abs}$) which have to be removed when measuring optical depth. 
In this subsection, we first present direct
measurements of \lyb absorption, then discuss the ratio of optical depths among
different transitions. \lyc measurements are discussed at the end of this subsection.

\begin{figure}
\epsscale{1.00}
\plotone{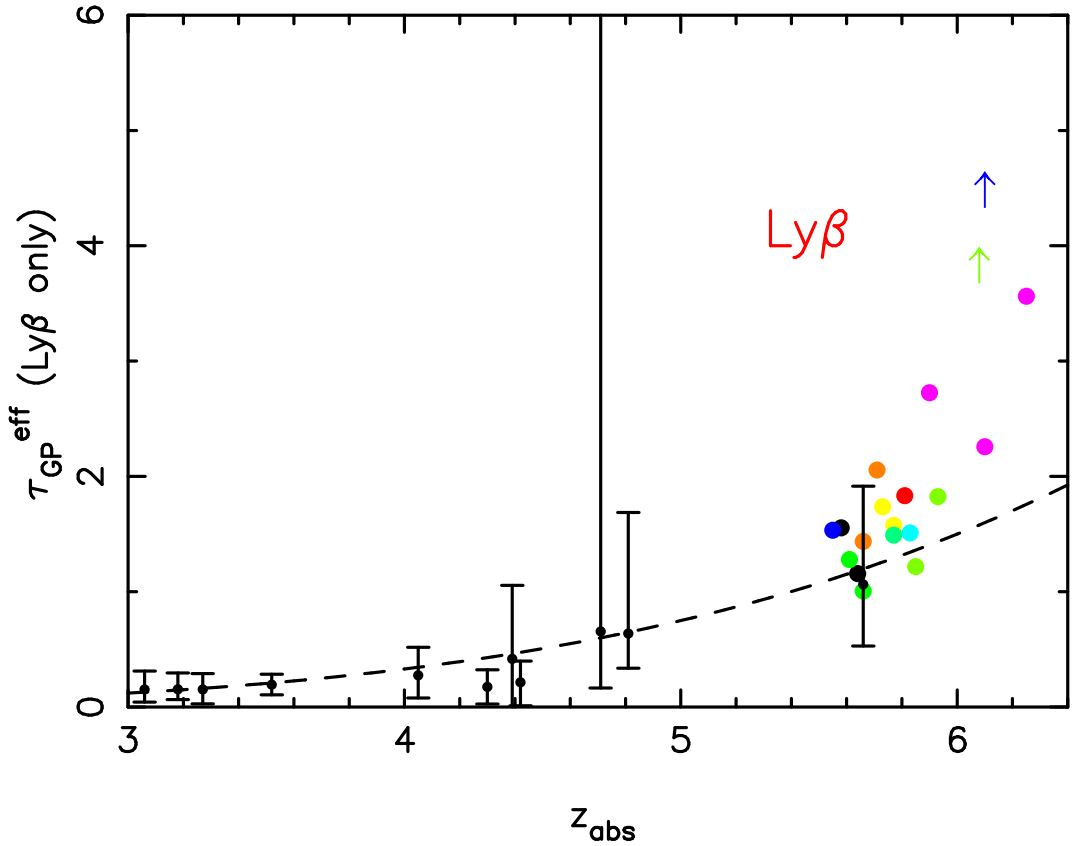}
\caption{Evolution of the Ly$\beta$ Gunn-Peterson optical depth with redshift.
At $z_{\rm abs} = 5.3 - 6.3$, there are 
19 independent measurements covering a total redshift interval
of $\Delta z = 2.9$.  These measurements have been corrected for the
foreground \lya absorption. We also plot the measurements at lower redshift from
Songaila (2004). The dashed curve shows the best-fit power-law for $z_{abs} < 5.5$:
$\tau_{\beta}^{\rm eff} = 0.38 \times \left(\frac{1+z}{5}\right)^{4.3}$. 
Like \lya, at $z_{\rm abs} \gtrsim 5.7$,
the evolution accelerates with a rapid deviation from the power-law relation.}
\end{figure}

We measure the \lyb transmission using the same procedure as for Ly$\alpha$.
The minimum redshifts for the \lyb measurements are chosen as:
$1+z_{\rm min}^{\beta} = (1+z_{\rm em})\times \rm 970\AA/\lambda_{\beta}$
to avoid contamination from \lyc absorption.
Because of this, 15 of the 19 lines of sight have only one independent 
measurement at $\Delta z = 0.15$. 
Two lines of sight have two independent data points, while 
two others have an uncontaminated \lyb redshift 
interval narrower than 0.15.
The \lyb measurements cover a redshift range of $5.3 < z_{\rm abs} < 6.3$
with a total redshift interval of 2.9.
Table 3 presents the observed transmitted flux ratio in the \lyb region,
without correcting for the foreground \lya absorption.
Figure 3 presents the GP optical depth due to \lyb absorption along the nineteen
lines of sight in our sample, as well as lower redshift measurements from
Songaila (2004). In Figure 3, we have corrected for the foreground
\lya absorption at $z'_{\rm abs} = (1+z_{\rm abs})\lambda_{\beta}/\lambda_{\alpha} -1$,
using Eq.~(5) above. 
Note that while this correction is statistically correct, it increases 
the dispersion of the \lyb GP optical depth.
The dashed line in Figure 3 is the best-fit relation at $z_{abs} < 5.5$: 
\begin{equation}
\tau_{\beta}^{\rm eff} = (0.38\pm0.05) \times \left( \frac{1+z}{5} \right)^{4.3\pm0.3}.
\end{equation}
We fixed the power-law index to be the same as Eq.~(5) in the fit.
Compared to Figure 2, the deviation from the power-law relation at $z_{abs} > 5.7$
is more pronounced in term of
the ratio of optical depths: at $z_{\rm abs} \sim 6$, the {\em lower limit} 
on the average \lyb optical depth is $\sim 3.5$, compared to the extrapolated
value of 1.7 from Eq.~(6). Also note that among the three \lya lower limits
in Figure 2 that also have valid \lyb measurements, one has a clear detection
of flux in \lyb, indicating that \lyb is indeed more sensitive to
large neutral fraction than Ly$\alpha$. The two other \lyb points,
however, still do not have detectable flux, suggesting a high neutral fraction.

A number of authors have studied the evolution of the ionizing background
and neutral fraction of the IGM by combining \lya and  \lyb measurements
(Paper I, Cen \& McDonald 2002, Lidz et al. 2002).
They show that the \lyb measurements indeed yield stronger constraints
on the ionizing background at $z_{\rm abs} > 6$. Songaila (2004) and
Oh \& Furlanetto (2005) explicitly estimated the ratio of effective optical
depths in the two transitions in a clumpy IGM. 
Here we expand the discussion in Paper I, and calculate the ratios using two
independent methods. Note that in Paper I, the difference between the \lya and \lyb
wavelengths was neglected in the calculations, resulting in an estimated
ionizing background that was 7\% {\em too high} at $z_{\rm abs} > 6$ (see \S4.1).

We first estimate the conversion factor using a model in which the IGM is isothermal and 
photo-ionized by an uniform UV background, 
following Paper I and similar treatments by other authors.
A detailed discussion of the formalism can be found in Paper I, and will
only be described briefly here.
We define the fractional density of the IGM as
$\Delta \equiv \rho/\langle \rho \rangle$.
Assuming local photoionization equilibrium, whereby the netural hydrogen
fraction depends on the square of the local density, one can show that
for a region of IGM with density $\Delta$,
\begin{equation}
\tau(\Delta) \propto \frac{(1+z)^{4.5} (\Omega_b h^2)^2 \alpha[T(\Delta)]}{ h\Gamma(\Delta,z) \Omega_m^{0.5}} \Delta^2, 
\end{equation}
where $\Gamma$ is the photoionization rate, assumed to
be uniform here, and $\alpha(T)$ is the
recombination coefficient at temperature $T$ (Abel et~al. 1997),
$\alpha(T) = 4.2\times10^{-13} (T/10^4\rm K)^{-0.7}
{\rm cm}^{3} {\rm s}^{-1}.$
The relation $T(\Delta) \propto \Delta^{\gamma}$ is the equation of state
of the IGM, where $\gamma = 0 - 1$. We assume $\gamma=0$ for isothermal case here.
The observed transmitted flux ratio or effective optical depth is averaged over
the entire IGM density distribution,
\begin{equation}
{\cal T} = \langle e^{-\tau} \rangle 
= \int_0^\infty e^{-\tau(\Delta)} p(\Delta) d(\Delta),
\end{equation}
where $p(\Delta)$ is the distribution function of the density of the
IGM.
We calculate ${\cal T}$ in a clumpy IGM using a parametric form for
the volume-weighted density distribution function $p(\Delta)$
(Miralda-Escud\'{e}, Haehnelt, \& Rees 2000):
\begin{equation}
p(\Delta) = A \exp \left[ - \frac{(\Delta^{-2/3} - C_0)^2}
                {2(2\delta_0/3)^2}\right] \Delta^{-\beta},
\end{equation}
where $\delta_0 = 7.61/(1+z)$, and $\beta$, $C_0$ and $A$ are numerical
constants given in Table 1 of Miralda-Escud\'e et al. (2000)
at several redshifts, derived from their simulation which assumed a 
cosmology consistent with that of 
WMAP. We linearly interpolate these constants  as a function
of redshift.
Eq.~(7) is appropriate for all Lyman series lines, with different
values of the proportionality constant.
Eq.~(7) also shows that if the ionizing background and the 
clumpiness of the IGM do not evolve with redshift, then
$\tau \propto (1+z)^{4.5}$. At $z_{\rm abs} < 5.5$, this is
the leading term in the observed scaling relation (Eqs.~5 and 6),
and the increasing clumpiness and decreasing ionizing background 
as one moves to  lower redshift roughly cancel each other out.
A significant deviation from this relation (\S3.3) indicates an
accelerated evolution in the ionizing background.

We calculate the GP optical depth in different Lyman transitions as a function
of IGM photoionization rate at $z = 6$.
Figure 4 shows the conversion factors $\tau^{\rm eff}_{\alpha} / \tau^{\rm eff}_{\beta}$
and $\tau^{\rm eff}_{\alpha} / \tau^{\rm eff}_{\gamma}$ as a function
of the photoionization rate $\Gamma_{-12}$, in units of
$10^{-12}$ s$^{-1}$. 
The conversion factor shows a mild dependence on the ionizing background.
For parameters of interest (see \S4.1), $\Gamma_{-12} = 0.01 - 0.25$,
the \lya - \lyb conversion factor $\tau^{\rm eff}_{\alpha} / \tau^{\rm eff}_{\beta} = 2.5 - 2.9$, and the \lya - \lyc conversion factor $\tau^{\rm eff}_{\alpha} / \tau^{\rm eff}_{\gamma} = 4.4 - 5.7$. These values are factors of $2-4$ smaller than those in the case
of a uniform IGM, consistent with those found by Oh \& Furlanetto (2005).

\begin{figure}
\epsscale{1.00}
\plotone{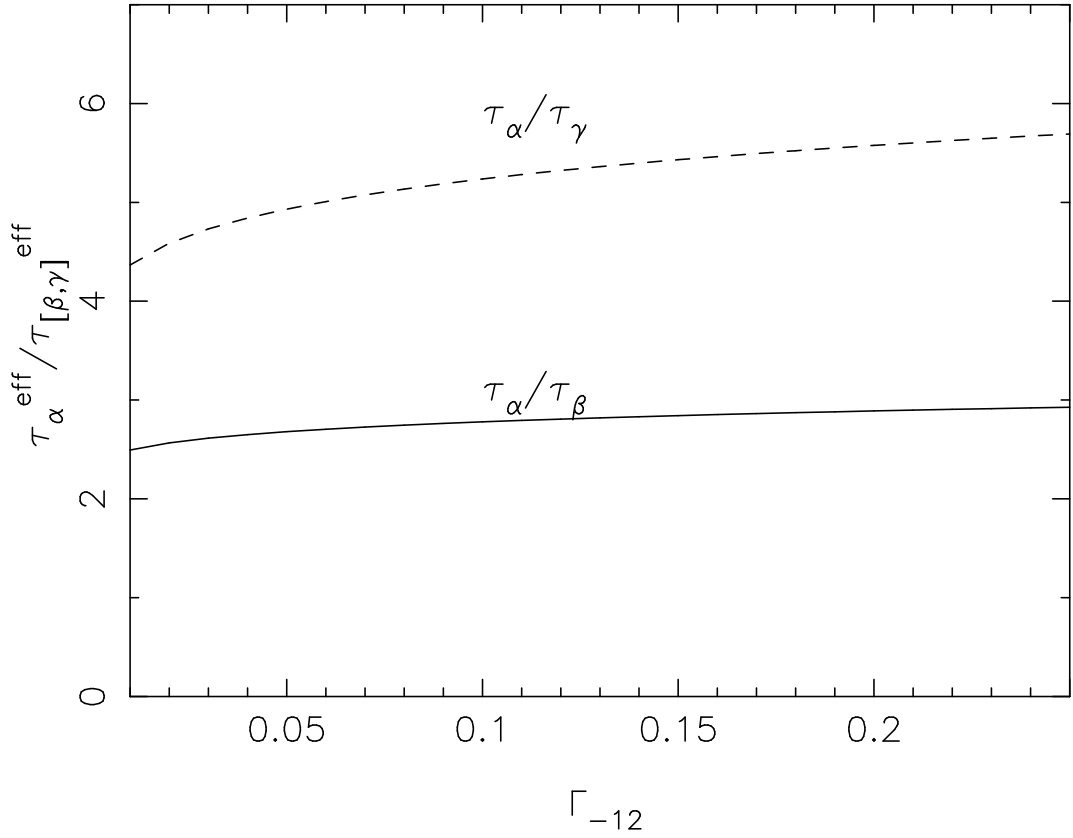}
\caption{The optical depth conversion factors between Lyman series transitions
as a function of the UV background at $z=6$. For a clumpy, isothermal IGM and a uniform
background, the  \lya - \lyb conversion factor lies in the range of 2.5 -- 2.9,
while the \lya - \lyc conversion factor lies between 4.4 and 5.7.}
\end{figure}

The calculation above is subject to the assumptions we made about
the clumpiness of the IGM, its equation of state, and the uniformity of
the ionizing background. We also calibrate the \lya - \lyb conversion
factor empirically: we calculate the average {\em transmitted flux ratios} 
(not the average {\em optical depth}) in \lya and \lyb over the same redshift range, $\cal T_{\alpha}$ and $\cal T_{\beta}$, the latter corrected
for foreground \lya\ absorption, 
and define the conversion factor as:
$\tau^{\rm eff}_{\alpha} / \tau^{\rm eff}_{\beta} = \ln({\cal T}_{\beta}/{\cal T}_{\alpha})$.
This method simply requires that the \lya and \lyb measurements yield consistent
results on the optical depth and neutral fraction. 
We find that for $z_{abs} = 5.4 - 5.8$: $\tau^{\rm eff}_{\alpha} / \tau^{\rm eff}_{\beta} = 2.25$,
and for $z_{\rm abs} > 5.9$, $\tau^{\rm eff}_{\alpha} / \tau^{\rm eff}_{\beta} = 2.19$.
This conversion factor is somewhat smaller than that found
using the photo-ionization model. Oh \& Furlanetto (2005) show that a more clumpy IGM,
or a UV background in which  $\Gamma$ is correlated with density, results in
a smaller conversion factor.
For the remainder of this paper, we adopt the conversion factors:
$\tau^{\rm eff}_{\alpha} / \tau^{\rm eff}_{\beta} = 2.25$, and $\tau^{\rm eff}_{\alpha} / \tau^{\rm eff}_{\gamma} = 4.4$. 
We will use the notation
$\tau^{\rm eff}_{\rm GP}$ as the effective optical depth converted to the \lya transition unless otherwise
noted.

The measurement of \lyc optical depth is severely restricted by the
presence of \lyd absorption at the low-redshift end and the quasar
proximity effect at the high-redshift end. Only the three lines of sight towards
the three highest redshift quasars at $z>6.2$ provide a redshift range 
$\Delta z > 0.06$ for this measurement. 
As before, the \lyc optical depth has been corrected for
foreground absorption in \lya and \lyb, using Eqs.~(5) and (6) and
converted to the \lya\ values.
The results are summarized in Table 4. In all three cases, the
\lyb regions probed correspond to redshift ranges with deep \lya GP troughs.

\begin{figure}
\epsscale{1.00}
\plotone{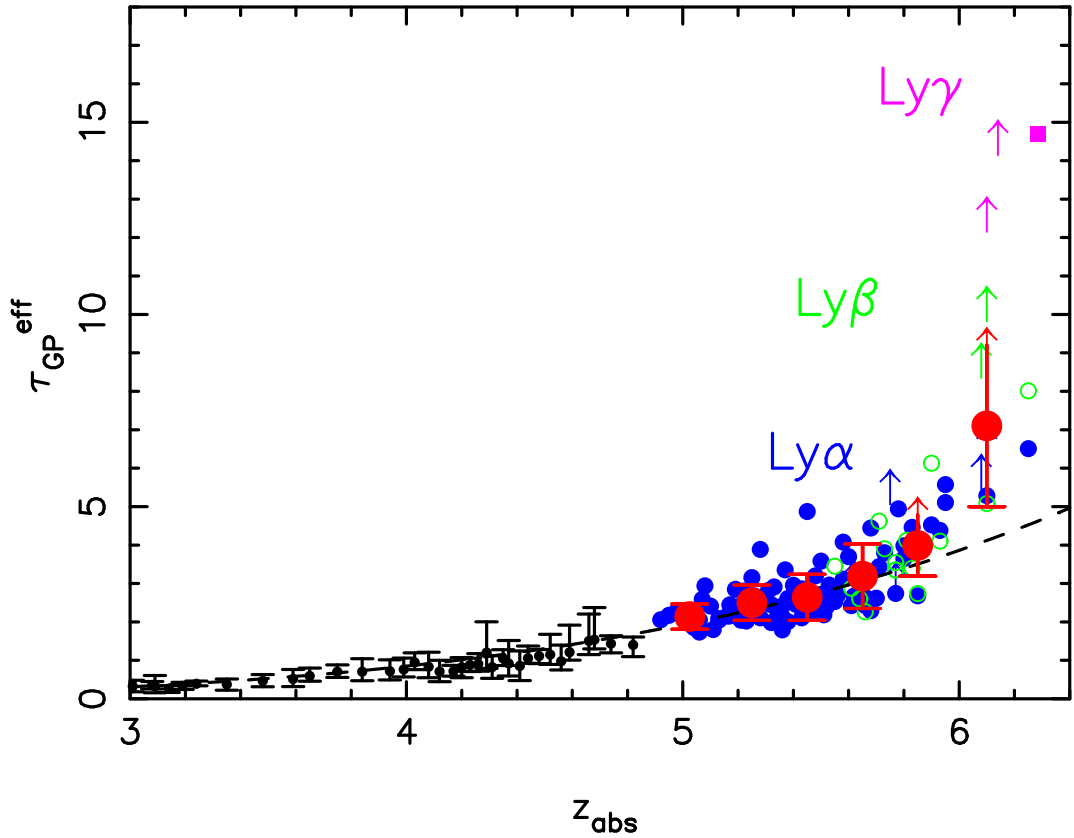}
\caption{Evolution of optical depth with combined the Ly$\alpha$, Ly$\beta$ and  \lyc results.
The \lyb measurements are converted to \lya GP optical depth using a conversion
factor of 2.25. 
The values in the  two highest redshift bins are lower limits since they both
contain complete GP troughs.
The dashed line is a redshift evolution of
$\tau^{\em eff}_{\rm GP} \propto (1+z)^{4.3}$. At $z>5.5$, the best fit evolution has
$\tau^{\em eff}_{\rm GP} \propto (1+z)^{>10.9}$, indicating an accelerated evolution.
The large filled symbols with error bars are the average and standard deviation
of optical depth at each redshift. The sample variance also increases
rapidly with redshift.
We also include the \lyc measurements (square symbol, Table 4); they cover only the high
redshift end of the \lya and \lyb bins due to contamination  from 
\lyd absorption along the line of sight at lower redshift.}
\end{figure}

\subsection{Accelerated Evolution of the Average and Scatter of GP Optical Depth}

Figure 5 shows the combined measurements of optical depth evolution using
the Ly$\alpha$, \lyb and \lyc transitions, with \lyb and \lyc measurements converted to their
\lya equivalent assuming conversion factors of 2.25 and 4.4, respectively.
For $z_{\rm abs} = 4.9 - 6.3$, we have also calculated the mean and
standard deviations of $\tau^{\rm eff}_{\rm GP}$ among different lines of sight at a redshift interval of 0.2 based on \lya and \lyb measurements.  
When calculating the average and standard deviation,
for points where both \lya and \lyb measurements are available, 
we use the \lya measurements in the cases where there is detected \lya transmitted
flux, and use the \lyb measurements or lower limits when no \lya flux is detected, regardless whether
\lyb flux is detected.  
When there is a non-detection, we simply use the 2-$\sigma$ lower limit values to calculate
the average and standard deviation. Therefore, they are lower limits 
for the two highest redshift bins. The results are tabulated in Table 5.
We do not use \lyc measurements in Figure 5 to calculate the average and
standard deviation, because they only cover a small redshift range. 
Including \lyc measurements increases the values of both the average and
the scatter of the GP optical depth. 

From the figure, we note that:

(1) 
At $z<6$, the \lya and \lyb measurements are fully consistent.
This is a result of  the calibration procedure used above.
At the highest redshift when no \lya flux is detected,
the \lyb optical depth measurement either gives a secure optical depth measurement
that is somewhat higher than the \lya lower limit (in one case), or
provides a much more stringent lower limit (in two cases).
Note that the \lyc measurements only cover the high redshift end of each bin
at which the absorption is particularly strong.

(2) With the additional constraints from \lyb measurements, the trend of accelerated
evolution in optical depth is clearly established: 
at $z_{abs} = 5.85$, the {\em lower limit} on the averaged optical depths
is 4.0, compared to the extrapolated value of 3.3 based on Eq.~(5);
and at $z_{abs} = 6.1$, the {\em lower limit} on the averaged optical depths is
7.1, compared to the extrapolated value of 3.8.
We fit a power-law for redshifts $z_{abs} > 5.5$, and find that 
\begin{equation}
\tau^{\em eff}_{\rm GP} \propto (1+z)^{>10.9}, \hspace{0.3cm} z_{\rm abs} = 5.5 - 6.3.
\end{equation}
This should be compared with $\tau_{\rm GP} \propto (1+z)^{4.3}$ at lower redshift, 
indicating a rapid change in the evolution of optical depth.

(3) The accelerated evolution of the optical depth  is accompanied by an
increase in the {\em dispersion}
of the optical depth: at $z_{\rm abs} < 5.0$, the standard deviation 
$\sigma(\tau) \sim 0.3$. It increases to $\sigma(\tau) \sim 0.6$ at $z_{abs} = 5.5$,
and  $\sigma(\tau) \sim 0.8$ at $z_{abs} = 5.8$.
At $z_{abs} > 6.0$, the lower limit on $\sigma(\tau)$ is 2.1.
The increased dispersion  $z_{abs} > 5.7$ is due to the fact that 
some lines of sight have complete GP troughs, with others have optical
depths only mildly higher than the power-law extrapolation from low redshift.
Different data points are independent measurements of the optical depth
in different IGM regions (\S4.1). This result suggests that the ionization state of
the IGM has large sample variance at the end of reionization (\S5).

Using a log-normal density distribution, and carrying out the
integral of Eqs.~(7), (8), and (9), Songaila \& Cowie (2002)
found:
\begin{equation}
\tau^{\rm eff}_{\rm GP} = A^{1/(1+3\beta/4)}-\frac{0.83}{\beta+4/3}\ln(A) + \rm const,
\end{equation}
where $A\propto 1/\Gamma$, and $\beta = 2-0.7\gamma$ for an uniform radiation
background (see also Oh \& Furlanetto 2005).
If the IGM is isothermal ($\beta = 2$), we have $\tau^{\rm eff} \propto \Gamma^{-0.4}$ as the leading term.
This scaling is {\em not} sensitive to the detailed IGM density distribution (Eq.~9).
Therefore, the evolution of the GP optical depth is driven by
the rate of evolution in the UV background:
if we observe that the GP optical depth increases by $\sim 2.5$ from
$z\sim 5.5$ to $6.2$, we can infer that the UV background decreases roughly by an order of magnitude.
We carry out a detailed calculation in the next section.

\section{Evolution of the Ionization State of the IGM}

In Paper I, we used the evolution of the GP optical depth and
the formalism described in Eqs.~(7) - (9) to
calculate a number of IGM ionization parameters, including
the ionizing background and neutral fraction
of the IGM. 
We also calculated the mean free path of ionizing photons using
the model of Miralda-Escud\'{e}, Haehnelt, \& Rees (2000).
Here we carry out similar calculations with the new dataset, and introduce
an alternative way of deriving the mean free path. 
However, we caution that the ionization parameters derived in this
section are model dependent:
they assume a particular model of the IGM density distribution
and photoionization equilibrium with a uniform ionizing
background.
We will discuss the limitations of these derived properties in detail
in subsequent subsections.
\subsection{Ionizing Background}

Figure 6 shows the estimated photoionization rate evolution with redshift 
based on Eqs.~(7) to (9) and the GP optical depth measurements in Figure 5.
Details of these calculations are discussed by \cite{MM01} and in Paper I.
 Consistent with the trend seen in Paper I,
the drop of ionizing background accelerates at $z_{abs} > 5.7$:
from $z_{abs} = 5.6 $ to 6.1, $\Gamma$ declines by a factor of $>4$.
In the deepest GP troughs, the 2-$\sigma$ upper limit of the
photoionizing rate derived from the \lyb
transition is  $\Gamma_{-12} < 0.025$. 
The most stringent constraint comes from the \lyc measurement (Table 4)
in quasar J1030+0524 at $z_{abs} = 6.11 - 6.17$, where $\tau_{GP} > 14.2$ (2-$\sigma$)
is consistent with a photoionizing rate  $\Gamma_{-12} < 0.012$ in that
region of the IGM, more than
an order of magnitude lower than the average background at $z_{abs} =5.5$.
Also note the relatively flat evolution between $z_{abs} = 5$ and 5.5.
This lack of evolution in $\Gamma$ was first noted by Cen \& McDonald (2002)
in a small sample of SDSS quasars. 
The IGM ionizing background is expected to increase
towards lower redshift as the  emissivity from
star forming galaxies and AGNs increases  and the  photon mean free path increases.
Cen \& McDonald (2002) suggested that the flattening might be due to the
change in the star formation rate as the
temperature and Jeans mass rises shortly after reionization.

\begin{figure}
\epsscale{1.00}
\plotone{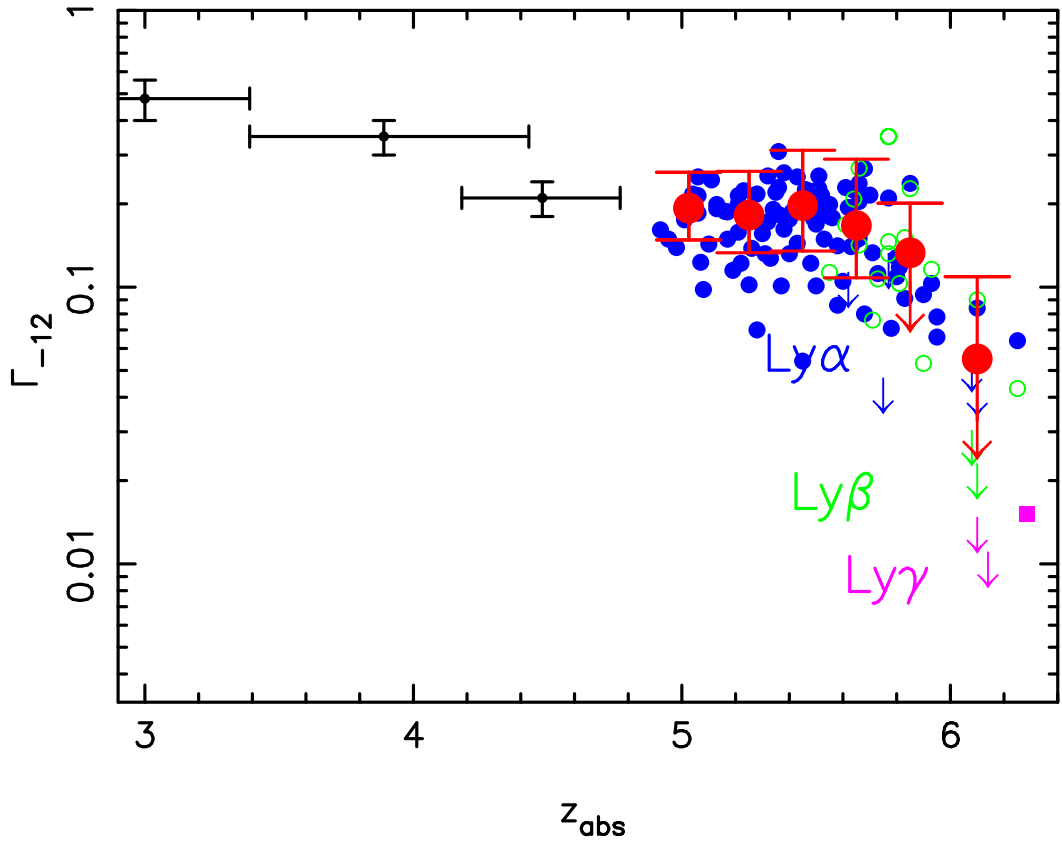}
\caption{Evolution of the photoionization rate (in units of
$10^{-12}\ {\rm s}^{-1}$) with redshift.
The points at $z_{\rm abs} < 4.8$ are taken from McDonald \& Miralda-Escud\'{e} (2001).
For the highest redshift points, measurements using
\lya (closed symbols) and \lyb (open symbols) transitions are shown separately.
Large, solid symbols with error bars are the average and 1-$\sigma$ dispersion
of $\Gamma$ at high redshift. As in Figure 5, we show the
\lyc measurements, but do not include them in the calculation of 
average and dispersion in the highest redshift bin.}
\end{figure}

Just as we saw in Figure 5, there is also a rapid increase in
the dispersion of the estimates of the ionizing background. 
At $z\sim 5$, the relative dispersion
$\sigma(\Gamma)/\Gamma$ is $\sim 30\%$.
It increases to $\sim 50\%$ at $z\sim 5.5$. 
At $z>6$, $\sigma(\Gamma)/\Gamma > 100$\% with $\Gamma_{-12}$
measurements ranging from $<0.02$ to $\sim 0.09$, 
reflecting the fact that some lines of sight have complete GP troughs
while others still have detectable transmission.
In the photoionization model, the observed fluctuation could be due to
two effects:
(1) large scale fluctuations in the underlying density field 
due to the finite length of the line of sight used in the measurement,
or (2) intrinsic fluctuations in the UV background and/or temperature of the IGM.

Data at $z_{\rm abs}>4.8$ in Figure 6 are derived from an effective optical
depth over $\Delta z = 0.15$ (\S4.1), corresponding to 44h$^{-1}$ comoving
Mpc at $z\sim 6$. The {\em one-dimensional} rms fluctuation of
the average mass density along a line of sight with length $L$ is given by:
\begin{equation}
\sigma^2_m(L) = \frac{1}{2\pi}\int^{\infty}_{0} dk \langle \tilde{F_k^2} \rangle \tilde{W}^2(kR),
\end{equation}
where $\tilde{W}(x) \equiv \sin x/x$ is the top-hat window function,
and $\langle \tilde{F_k^2} \rangle$ is the one-dimensional power spectrum,
 related to the normal three-dimensional linear power spectrum as (\cite{KP91}):
\begin{equation}
\langle \tilde{F_k^2} \rangle = \frac{1}{2\pi}\int_{k}^{\infty} P(y) y dy.
\end{equation}

Using the linear power spectrum determined from the \lya forest
at $z\sim 3$ by McDonald et al. (2005), and assuming that it grows linearly
with scale factor at high redshift, we find that at $z\sim 6$, the one-dimensional
r.m.s. density fluctuation is $\sigma_m(L) = 0.11$
over $\Delta z = 0.15$.
From Eq.~(7), $\Gamma \propto \rho_b^2$ for a given optical depth; a 10\% fluctuation in the  average density leads to
a fluctuation of $\sim$ 20\% in the resulting $\Gamma$ value.
This value is somewhat smaller than the dispersion of the $\Gamma$ measurements at $z\sim 5$,
and is much smaller than that at higher redshift.
Over such a narrow redshift range, the average IGM temperature is not likely
to have changed by more than a factor of two (Hui \& Haiman 2003),
resulting in a difference in estimated $\Gamma$ by $<60\%$, since
$\tau \propto \alpha(T) / \Gamma \propto T^{-0.7} / \Gamma$ (\S3.2).
Therefore, we interpret the large dispersion of optical depth and derived 
ionizing background at $z>5.7$ as a result of fluctuations in the
UV background near the end of reionization: even over the scale size of
44 h$^{-1}$ comoving Mpc, the average UV background fluctuates by a factor
of $\gtrsim 4$.
The dispersion of estimated $\Gamma$ values at $z<4$ is much smaller
and is mostly due to sample variance of the average density
along the line of sight.

Large fluctuations in the UV background 
also imply that the transmission is correlated over large scales.
We calculate the line of sight dispersion of transmitted flux by 
comparing the average transmitted flux over neighboring bins along the
same line of sight, and
subtract from it the difference due to redshift evolution (Eq.~5).
We find that at $z\gtrsim 5.5$, when averaging over redshift
interval $\Delta z < 0.1$, the dispersion of line of sight of GP optical depth 
is $\sim 20$\% {\em smaller} than the dispersion when comparing different
lines of sight. 
Thus the GP optical depth is correlated over scales 
of tens of comoving Mpc.
At $\Delta z > 0.15$, this difference is $<10$\%. 
At larger $\Delta z$, the test becomes difficult 
due to the limited line of sight towards any particular quasar.
Because of the limited sample size, non-uniform S/N of the sample spectra,
and strong redshift evolution, we do not attempt to
calculate the transmission two-point correlation function in this paper
(e.g. McDonald et al. 2000). 
By observing high-redshift quasar pairs that have projected separations 
of tens of comoving Mpc, one could check this correlation in the
transverse direction (e.g. Mahabal et al. 2005).

Large fluctuations in the UV background are a generic feature of
the IGM close to the end of reionization, when the individual HII regions
start to percolate and most of the UV photons come
from the earliest highly clustered star forming galaxies.
Note that in our photoionization model, we assumed a uniform UV
background. Here we show that at the end of reionization, there is
a large background fluctuation.
The derived UV background fluctuations will provide a sensitive
test of models of the reionization history when compared directly with simulations.

\subsection{Neutral Fraction and Mean Free Path of the IGM}

We use the GP optical depth measured in \S3 and the IGM density distribution,
Eq.~(9), 
to calculate both the volume- and mass-averaged neutral hydrogen fraction.
Figure 7 shows the
estimated volume-averaged neutral fraction using our sample of
nineteen quasars. 
It is compared with a recent simulation
by Gnedin (2004), which has a sharp transition in the  ionization state
at $z \sim 5.5 - 6.5$, and a small total Thompson scattering optical depth
to CMB $\tau_e = 0.06$ (compare to the WMAP three-year value of
$\tau_e = 0.09 \pm 0.03$, Spergel et al. 2006, see \S 4.3 for more discussion on
the model comparisons). The strong evolution of neutral fraction in Paper I
is confirmed here: there is an order of magnitude increase between
$z\sim 5.1$ and 6.1, with the evolution accelerated at $z>5.7$. 
We emphasize that the calculation of volume- or mass-averaged neutral fractions depends on the  density distribution
function of the IGM as well as on the assumption of a uniform UV background.
Oh \& Furlanetto (2005) show that most of the IGM
transmission is through the most highly ionized, underdense regions
(see also Figure 7 of Paper I).
 As discussed in Paper I, the volume averaged values
are less affected by the ionization state of high-density regions than are the mass averaged values. 
Over a narrow redshift range where the density 
distribution of the IGM evolves little, the {\em ratio} of neutral fractions
at different redshifts is a robust quantity, especially when
compared with simulations.

\begin{figure}
\epsscale{1.00}
\plotone{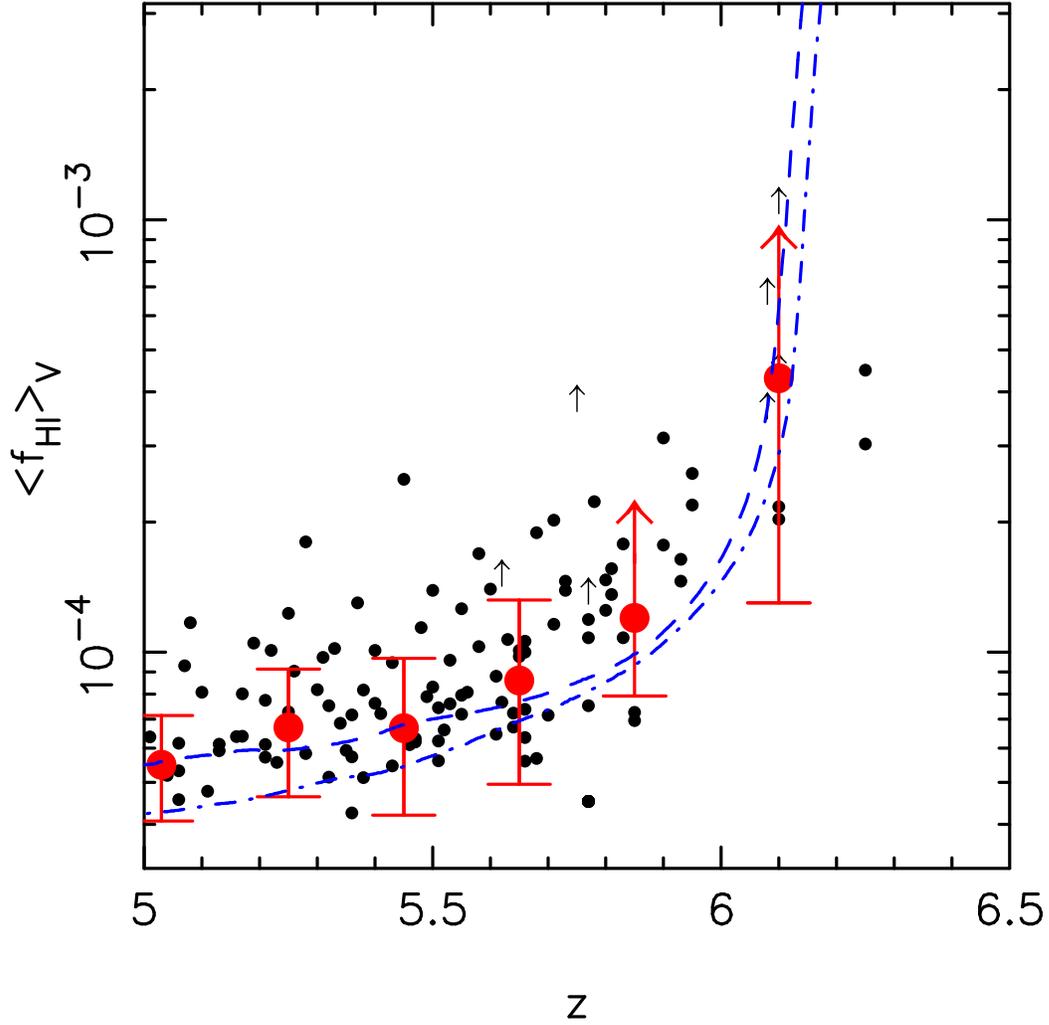}
\caption{Evolution of volume-averaged neutral hydrogen fraction of the IGM.
The small dots are measurements based on the
nineteen high-redshift quasars,
while the large points with error bars are the means in bins of redshift.
The dashed and dotted-dashed lines are the  volume-averaged
results from the simulation of Gnedin (2004) with  box sizes of 
4 h$^{-1}$ Mpc and 8 h$^{-1}$ Mpc, respectively.}
\end{figure}

\begin{figure}
\epsscale{1.00}
\plotone{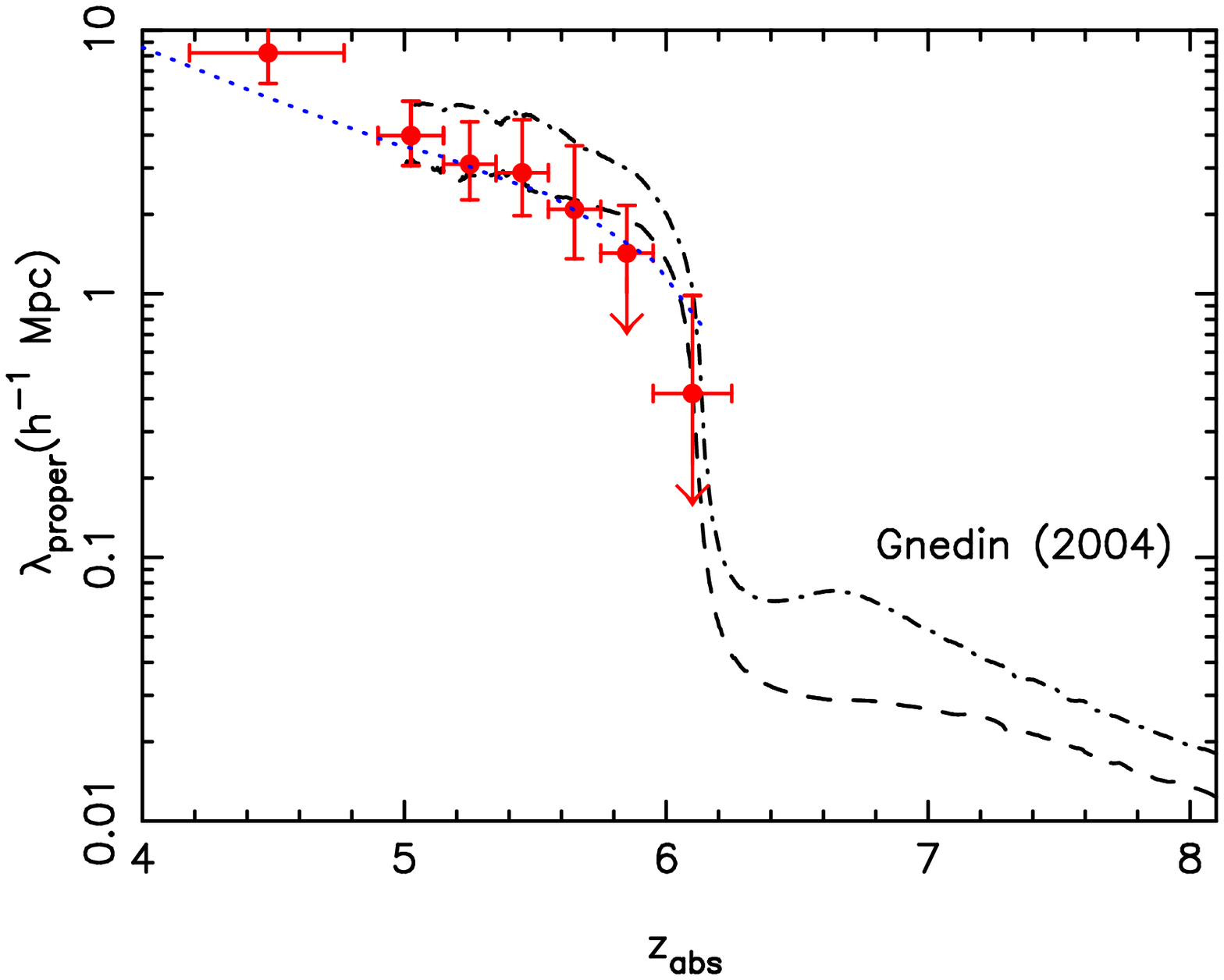}
\caption{Evolution of the mean free path of ionizing photons inferred
from the
observations using Eq.~(13) (solid symbols with error bars)
and using the method presented in Paper 1 (dotted line),
compared with simulations of Gnedin (2004, dashed and dashed-dotted-dashed lines).
The mean free path at $z\sim 5.7$ is comparable to or smaller than
the correlation length of high-redshift star forming galaxies.
}
\end{figure}

In Paper I, we calculated the evolution of the mean free path of
ionizing photons, using the model of Miralda-Escud\'{e}, Haehnelt, \& Rees
(2000). They model reionization using  the evolution of critical
overdensity $\Delta_i$, above which the IGM is still mostly neutral,
and below which it is mostly ionized. 
$\Delta_i$ increases as the reionization front progresses to higher density regions.
They also suggest an expression for the ionizing photon mean free path:
$\lambda_i = \lambda_0 [1-F_{\rm v}(\Delta_i)]^{-2/3}$,
where  $F_{\rm v}(\Delta_i)$ is the fraction of volume with
$\Delta < \Delta_i$, and
$\lambda_0 H$ = 60 km s$^{-1}$,
which assumes that the shapes of isolated density contours remain roughly constant
as $\Delta_i$ decreases..
However, the critical overdensity $\Delta_i$ is not well defined
in the simple photoionization picture; we assumed that 95\% of the total
neutral gas lies above $\Delta_i$ in Paper I. 
In addition, overdense regions in the IGM are likely to contain the earliest star-forming
galaxies that provide most of the ionizing photons, thus they might have been
ionized earlier.
For these reasons, in this paper, 
we propose an alternative approach to calculate the mean free path $\lambda$ by
simply using the mass averaged neutral hydrogen density:
\begin{equation}
\frac{1}{\lambda} = \langle \sigma_{\nu} \rangle \frac{\int_{0}^{\Delta_c}
n_{\rm HI}(\Delta) p(\Delta) d\Delta}{\int_{0}^{\Delta_c}p(\Delta) d\Delta},
\end{equation}
where $\langle \sigma_{\nu} \rangle$ is the frequency averaged cross section
of Lyman limit absorption, assuming an ionizing spectrum $J_v \propto \nu^{-5}$ (Barkana \& Loeb 2001),
and $n_{\rm HI}(\Delta)$ is the neutral fraction corresponding to a certain
overdensity calculated using Eqs.~(7) - (9).
The integrals over the density distribution are only
carried out to $\Delta_c = 150$, above which the region is already
virialized and is no longer part of the IGM. 
This expression is accurate in cases in which the mean free path is larger
than the scale of IGM clustering, or the UV photon field follows
the IGM density distribution; these assumptions might not be valid at the
early stage of reionization.
Furthermore, these calculations depend on the derived neutral fraction
and assumed IGM density distribution.
While the exact values of the mean free path are model-dependent, the
strong evolutionary trend is clearly detected.
Figure 8 shows the mean free path calculated using Eq.~(14), where both
the mean and 1-$\sigma$ error bars are presented.
The dotted line shows the mean free path calculated using the method in 
Miralda-Escud\'{e} et al. (2000) and in Paper I.
The two results are fully consistent. They again show the
accelerated transition at $z_{abs} >5.7$, where only an upper limit can be 
derived for complete GP trough regions. 

At $z>5.7$, the mean free path is smaller than 0.5 h$^{-1}$ proper Mpc 
(3.5 h$^{-1}$ comoving Mpc).
In comoving units, this is comparable to the correlation length of
galaxy clustering at $z\sim 0$, 
($r_0 \sim 5$ h$^{-1}$ Mpc, Zehavi et al. 2005), and of Lyman break galaxies at $z\sim 2 - 4$ 
($r_0  \sim 4$ h$^{-1}$ Mpc, \cite{adelberger05}),
and  at $z= 4 - 5$
($r_0 \sim 4 - 8$ h$^{-1}$ Mpc, \cite{kashikawa05}).
At $z >5 $, evidence of galaxy clustering on scales of $\sim 10$ Mpc are 
reported in Hu et al. (2005), and \cite{malhotra05}.
High-redshift star forming galaxies, which likely provide most of the 
UV photons for reionization, are highly biased and clustered at 
physical scales similar to the mean free path. The assumption 
of uniform UV background under which we have derived these results
is no longer valid. We saw clear evidence
for non-uniform $\Gamma$ in \S4.1.

\subsection{Comparison with Hydrodynamical Simulations}

Hydrodynamical simulations
(Gnedin 2002, 2004, Razoumov et al. 2002, \cite{pn05})
have been used to calculate the evolution of the GP optical depth during the
epoch of reionization.
Gnedin (2004) presented a SLH simulation of reionization by stellar sources
with  a new recipe for radiative transfer using an optically thin variable Eddington tensor
approximation and box sizes of 4 -- 8 h$^{-1}$
comoving Mpc. 
In these simulations, the mean mass-averaged neutral fraction
reaches $10^{-2}$ and 0.5 at $z=6.3$ and 7.3, respectively.
The reionization process is modelled as similar to a phase transition,
completed within $\Delta z \sim 1$.
Similarly, Paschos \& Norman (2005) presented a set of reionization simulations using the Eulerian
code, Enzo, and a box size of 6.8 h$^{-1}$ co-moving Mpc.
Their default model has a sharp reionization at $z_{\rm reion} \sim 7$,
around which the mass-average neutral fraction increases from 10$^{-3}$ to of
order unity.
Both simulations have been calibrated according to the average GP optical depth
measurements at $z<6$ using earlier
datasets (e.g. Becker et al. 2001, Songaila 2004), 
so the consistency in measurements in this redshift range is by design.

\begin{figure}
\epsscale{1.00}
\plotone{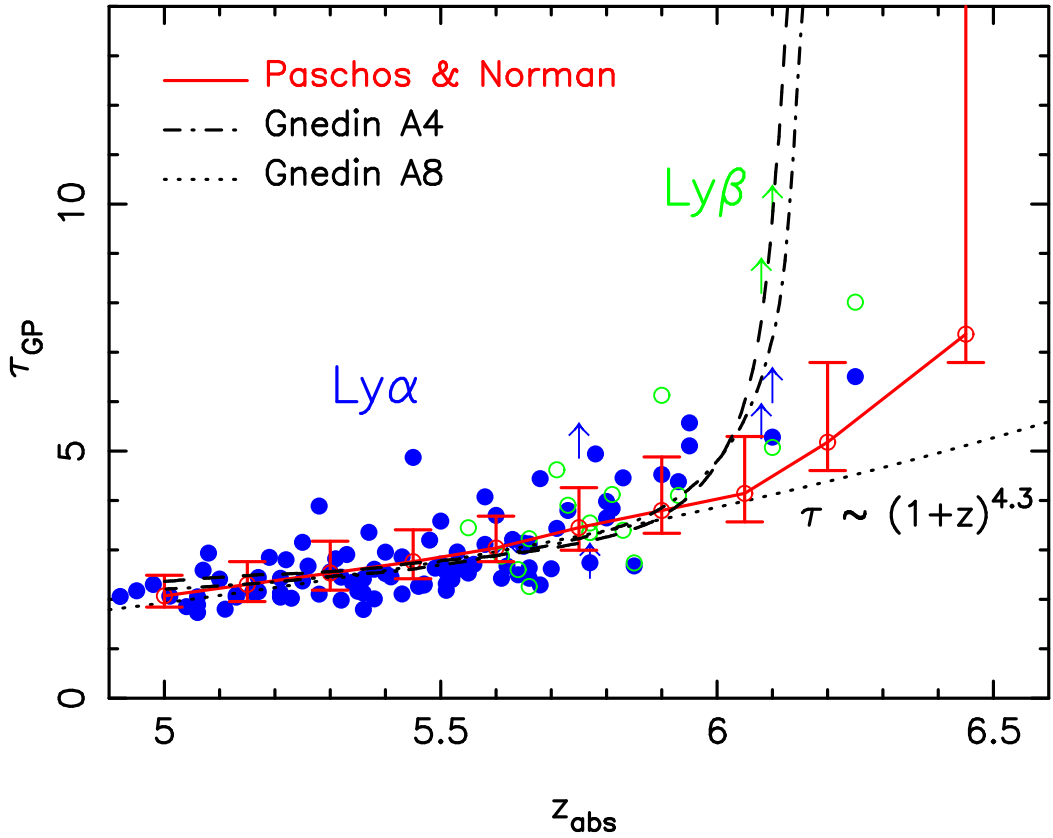}
\caption{Comparison of the evolution of observed GP optical depth with simulations.
Data points are the same as in Figure 5. The dotted line is the power-law
fit to optical depth evolution at $z<5.5$: $\tau_{\rm GP} \propto (1+z)^{4.3}$.
The solid line with error bars is the result from simulations of Paschos \& Norman
(2005), which have $z_{\rm reion} \sim  7$.
The dashed and dotted-dashed lines are the 4 Mpc and 8 Mpc simulations
of Gnedin (2004), which have  $z_{\rm reion}\sim 6.5$. The data points 
fall between the two simulations, but are somewhat closer to those of Gnedin.}
\end{figure}

However, these simulations which model the reionization process as 
a phase transition could be too simplistic for two reasons.
First, both simulations include only a simple prescription of the evolution of
emissivity of ionizing sources.
The simulations have not captured all the relevant physics,
especially the effects of early galactic feedback and large scale structure.
The total Thompson scattering optical depth ($\tau_e$) is less than  0.10
in both simulations, inconsistent with the WMAP first-year measurements at 
face value ($\tau = 0.17 \pm 0.06$, Kogut et al. 2003, Tegmark et al. 2004),
but is consistent with the WMAP three-year measurements (Spergel et al. 2006).
A number of semi-analytical analyses 
(e.g., Cen 2003a,b, Chiu et al. 2003, Wyithe \& Loeb 2003a, Haiman \& Holder 2003 etc.),
show that adding feedback from an early generation of galaxy formation, or
adjusting the star formation efficiency at $z>6 - 10$,
will produce a prolonged, or multi-epoch reionization history.
These models predict a GP optical depth consistent with quasar observations
at $z<6.5$, while having  a large total Thompson scattering optical depth.

The second problem is the impact of large scale structure.
Both simulations have a limited box size ($<10$ comoving Mpc).
However, Wyithe \& Loeb (2004) and Furlanetto \& Oh (2005) have studied
growth of ionized bubbles at the end of reionization,
and estimated that the bubble have sizes of $>10$ Mpc at the end of
the overlapping phase. 
Wyithe \& Loeb (2004, 2005) showed that the observed redshift of
overlap along different lines-of-sight has a variation 
of $\Delta z \sim 0.15$, and the GP optical depth
variations should be of order unity on a scale of $\sim 100$ comoving Mpc.
These results are consistent with the large scale variation of the UV background discussed
in \S4.1. 
Thus to  fully understand the reionization process, simulations with box sizes of
the order 100 Mpc are needed (e.g., Kohler et al. 2005).

Figure 9 compares the GP optical depth measurements of Figure 5 with
the simulations of Gnedin (2004) and of Paschos \& Norman (2005). We also compare the estimated
IGM neutral fraction and mean free path of UV photons with those
of the Gnedin simulations in Figures 7 and 8. 
In the simulations, the onset of reionization is characterized by 
an acceleration in the evolution of the GP optical depth. 
Paschos \& Norman (2005) showed that when $z_{\rm reion} - z > 1$, 
the GP optical depth scales as  a simple power-law
$\tau_{\rm GP} \propto (1+z)^{4.2}$, consistent with what we find at
$z_{abs} < 5.7$. For the last $\Delta z = 0.5 - 1$, 
as the epoch of reionization approaches, the optical depth
evolution accelerates, again mirroring the observed trend.
The observed points fall in the range covered by the three simulations.
A similar rapid change is noted in the volume-averaged
neutral fraction (Figure 7) and photon mean free path (Figure 8),
in which the simulations closely match the values estimated from observations. 

The simulations also show that the end of reionization is accompanied by 
a rapid increase
of dispersion in optical depth.
Paschos \& Norman (2005)  calculated the flux variance among different lines of sight
by observing through random lines of sight in the simulation box.
The solid error bars in Figure 9 are the 68\% range of the optical depth measurements
at each redshift bin. At $z<5.5$, the dispersion in the simulation
matches the observations well. It also reproduces the trend of increased
scatter at $z>6$, although at somewhat higher redshift.
This accelerated evolution corresponds to the stage of reionization
where individual HII regions begin to overlap and the IGM ionization
experienced a rapid transition (e.g. Gnedin 2002).
The comparison of the observations with simulations strongly suggest
that $z\sim 6$ corresponds to the end of the overlapping stage of reionization,
but for reasons mentioned earlier, it does not provide 
detailed guidance to what happened at much higher redshift,
when the reionization history may have been more complex.

\section{Distribution of Dark Gaps}

In the last three sections, we have used the evolution of the effective optical
depth to probe the ionization state of the IGM. 
The effective optical depth directly probes the neutral fraction
and UV ionizing background in photoionization models. However, it
is a low order statistic.
At $z>5$, the appearance of the \lya forest is the reverse of that at low redshift:
most regions along the line of sight are optically thick, while a small fraction
of transparent ``spikes'' provide most of the \lya transmission.
Meanwhile, ever longer dark absorption gaps  --  continuous regions
with large optical depth --  emerge at
$z_{abs} > 5.7$.
A successful theory of IGM evolution at the end of reionization
needs to reproduce not only the evolution of the mean optical depth,
but also the geometry and topology of IGM transmission and the dark gaps.
In Pentericci et al. (2002) and in Paper I, we examined the distribution
of transparent spikes along a few lines of sight.
In this section, we expand these previous studies and discuss three issues:
the presence of complete  GP absorption troughs at the highest redshift (\S5.1);
the evolution of the dark gap distribution (\S5.2); and 
the use of dark gap lengths to constrain the IGM neutral fraction (\S5.3).

\subsection{Gunn-Peterson Troughs in the Highest-Redshift Quasars}

The presence of complete Gunn-Peterson troughs is the most compelling
evidence that the IGM is approaching the end of the reionization epoch at 
$z\sim6$. 
To probe the evolution of the IGM at $z>6$ requires quasars at $z>6.1$ 
to not be affected by the quasar proximity effect.
Figure 11 shows the regions immediately blueward of the \lya and \lyb emission
lines for the five quasars in our sample at $z>6.1$;
deep absorption troughs with $\tau^{\rm eff}_{\rm GP} > 5$
and length $\Delta z > 0.1$ are indicated.
The ends of troughs are chosen as locations at which significant transmission
spikes  with $\tau < 2.5$ are detected.
Table 6 summarizes the GP optical depth measurements for these troughs.
We now briefly discuss the case for each quasar:

\begin{figure}
\epsscale{1.00}
\plotone{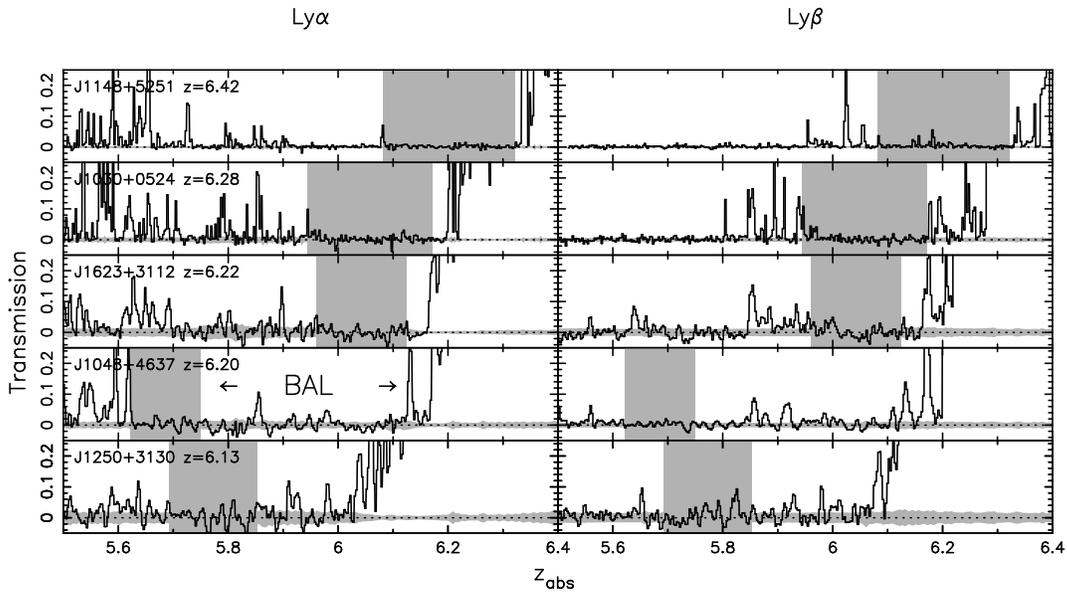}
\caption{Deep GP troughs in the five highest-redshift quasars known
at $z>6.1$, in both the \lya and \lyb transitions. 
The grey band around zero transmission represents the 1-$\sigma$
uncertainty of the data.
The ends of troughs are chosen as locations at which significant transmission
spikes are detected.
SDSS J1048+4637 is a BAL quasar, in which we use the redshift of
the CIV BAL feature as the upper bound of the GP trough.
These quasars show significant line of sight 
variation in the GP optical depth measurements and the redshift 
where deep troughs first appear.}
\end{figure}

\noindent
{\bf SDSS J1148+5251 ($z=6.42$)}. This object is discussed in
detail in White et al. (2003, 2005) and Oh \& Furlanetto (2005).
Extended dark regions in \lya absorption start at $z_{abs} > 5.91$;
however, the spectrum shows clear transmissions in the \lya - \lyc transitions at this redshift.
White et al. (2003) speculated that the \lyb transmission could be
due to \lya emission from intervening galaxies at $z\sim 4.9$,
as suggested by strong CIV absorption at that redshift.
Subsequent HST imaging indicated that the flux in the \lyb trough is a point
source coincident with the quasar position, inconsistent with the intervening
galaxy hypothesis.
The \lya and \lyb troughs yield consistent optical depth
measurements: $\tau^{\rm eff}_{\rm GP} \sim 5.5 - 7$. 
The consistency between the \lya and \lyb GP optical depth
measurements is direct evidence that at $z\sim 6$,
the IGM is already highly ionized at least along this line of sight,
although the neutral fraction has increased by a factor of $\sim 4$ 
from that at $z\sim 5.7$ even
in this most transparent line of sight among $z>6.1$ quasars. 
Table 4 also shows the \lyc measurements in the highest-redshift interval
at $z = 6.25 - 6.32$, where the transmission is detected at 3-$\sigma$
level, yielding an effective optical depth: $\tau^{\rm eff}_{\rm GP} \sim 15$.
No \lya or \lyb transmission is detectable over this narrow redshift range,
consistent with the \lyc measurement, and suggesting large fluctuations in 
the neutral fraction  along the line of sight.
Note that there is also a long \lya trough at $5.91 < z_{\rm abs} < 6.08$,
although it has clear transmission in \lyb, indicating a high level
of ionization.

\noindent
{\bf SDSS J1030+0524 ($z=6.28$)}. This is the first quasar at $z>6$ in which a
complete GP trough was detected (Becker et al. 2001); it still has the
longest and deepest absorption trough among known quasars, in which  no flux in any of the Lyman
transitions is detected, yielding the most stringent constraint on
the evolution of the neutral fraction. 

\noindent
{\bf SDSS J1623+3112 ($z=6.22$)}. This line of sight is very similar to that
of SDSS J1030+0524, with complete GP troughs in the \lya and \lyb transition
immediately blueward of the emission lines. The starting redshift
and length of the complete GP trough are also similar.
This object is fainter and has
a shorter total exposure time, yielding an upper limit 15\% weaker
than that of SDSS J1030+0524. It also has a \lyc trough with no detectable
transmission. 

\noindent
{\bf SDSS J1048+4637 ($z=6.20$)}. This is a BAL quasar (Fan et al. 2003,
Maiolino et al. 2004). The SiIV absorption extends to $z_{abs} \sim 5.83$,
and a somewhat weaker CIV trough reaches to $z_{\rm abs}=5.75$.
We therefore can only measure the IGM properties at lower redshift. 
The quasar has a dark trough at $z = 5.63 - 5.75$, where no flux
is detected in either \lya or \lyb at a level comparable to SDSS J1623+3112.
This is at a redshift considerably lower than the other lines of sight.
We regard it as less reliable, 
as the trough could be affected by weaker BAL features as well as by
absorption troughs from
other ions (e.g. PV$\lambda1121$, Arav et al. 2001).

\noindent
{\bf SDSS J1250+3130 ($z=6.13$)}. This object shows a dark \lya GP trough
at $z=5.69 - 5.85$ with an optical depth $\tau^{\rm eff}_{\rm GP} >4.6$ (2-$\sigma$). 
The trough doesn't start
immediately blueward of the \lya emission line, and has detectable \lyb transmission,
with $\tau^{\rm eff}_{\rm GP} \sim 6.5$. The \lyb measurement is comparable to
that of SDSS J1148+5251 at considerably higher redshift,
although  the S/N of the spectrum is lower for this quasar.

The properties of these absorption troughs show significant variation
between lines of sight. Two of the three quasars at $z>6.2$
show complete \lya - \lyc troughs, while the third one has only a somewhat higher
optical depth than seen at lower redshift. 
On the other hand, some dark troughs already start to appear at $z\sim 5.7$.
We quantify the evolution of the dark absorption gaps in the next subsection.

\subsection{Dark Gap Distribution}

Statistics of dark gaps (e.g. Croft 1998) were first applied to a sample
of $z = 4 - 6$ quasars by Songaila \& Cowie (2002). They studied the 
evolution of continuous absorption gaps with large optical depth
($\tau > 2.5$) and found that extended absorption troughs such as those observed
by Becker al. (2001) in SDSS J1030+0524 can be explained by large variations
in the UV background at $z>6$.
Barkana (2002) presented a semi-analytic model for the length of dark gaps
as a function of IGM HII region filling factor.
Furlanetto, Hernquist, \& Zaldarriaga (2004) also presented a model of dark gap
distribution, considering source clustering and the inhomogeneous 
IGM.
Paschos \& Norman (2005) and Kohler et al. (2005) studied the dark gap distribution using spectra
derived from their hydrodynamical simulations. 
They suggest using the distributions of dark gaps as a statistic to
compare with observations, and found that the onset of reionization
is associated with a dramatic increase both in the average and dispersion
of gap length. 

\begin{figure}
\epsscale{0.60}
\plotone{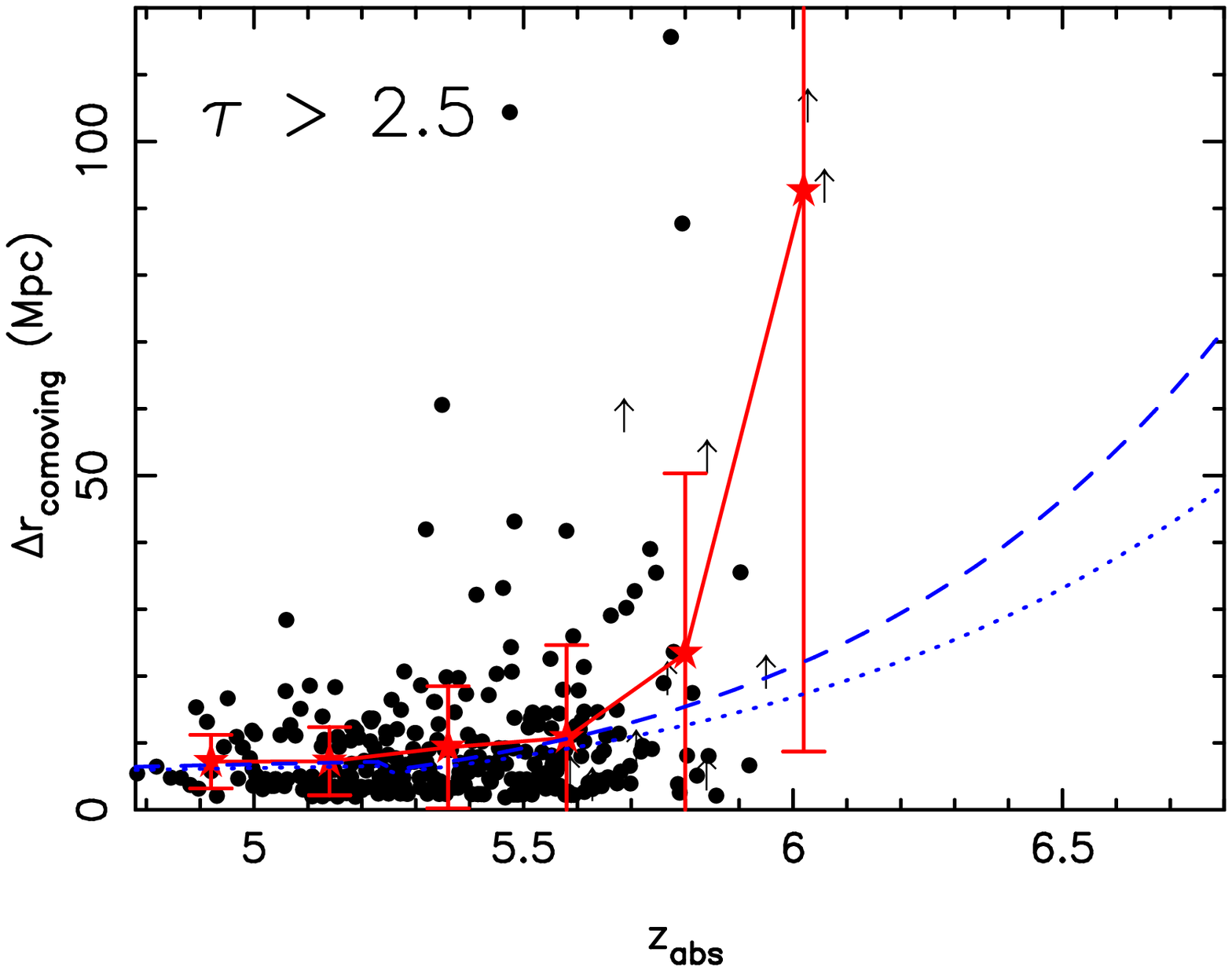}
\plotone{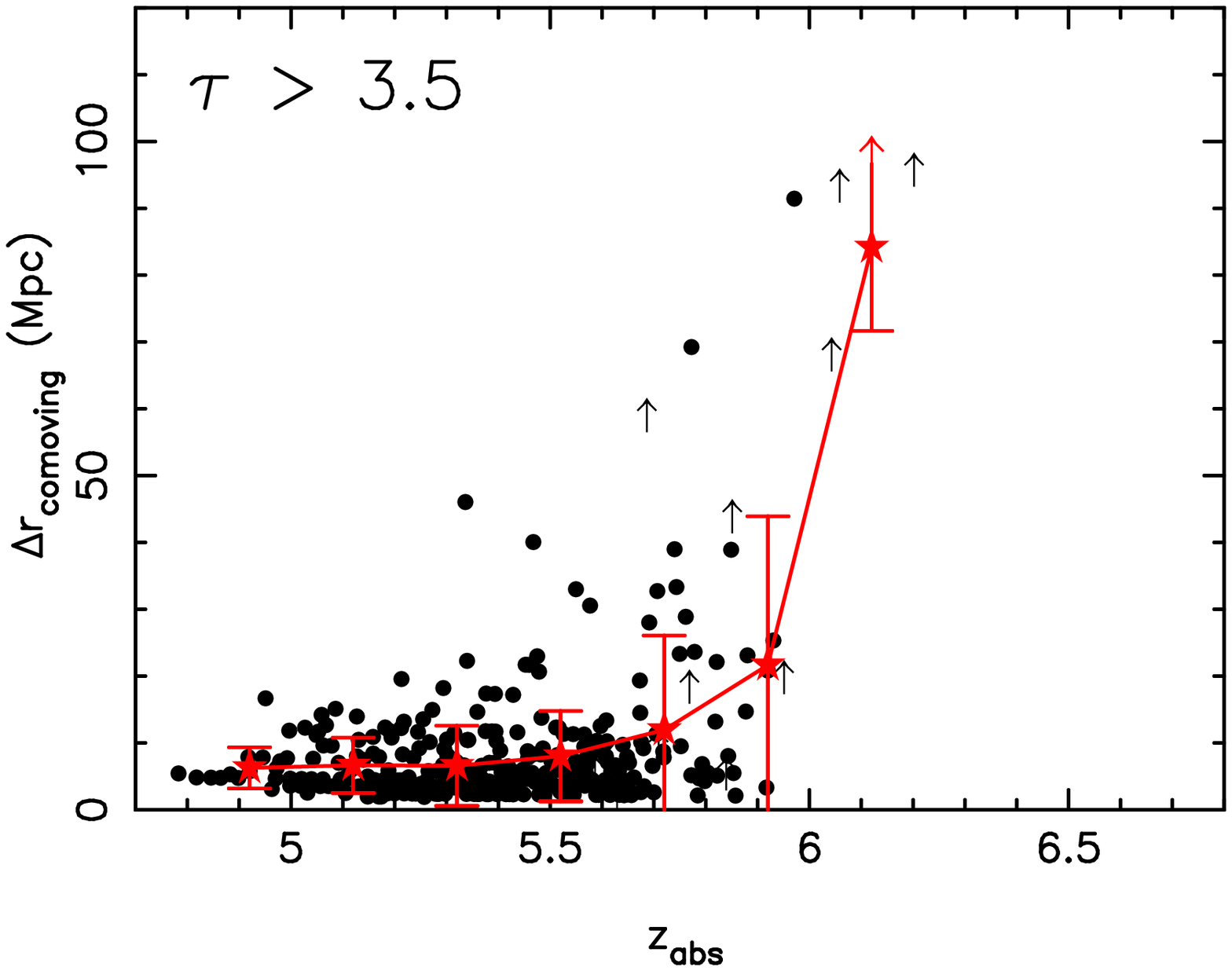}
\caption{Distributions of dark gap lengths, defined as continuous regions in the spectra in which 
all pixels have observed optical depth
larger than 2.5 (upper panel) and 3.5 (lower panel) for the \lya transition in
the twelve lines of
sight that have Keck/ESI observations.
Upward arrows are gaps immediately blueward of the quasar proximity zone, therefore
the length is only a lower limit.
Solid lines with error bars are
mean depth lengths, with 1-$\sigma$ dispersion given at each redshift bin.
Long dark gaps start to appear at $z\sim 5.6$, and  the average gap length
increases rapidly at $z>5.9$.
In the upper panel, we also compare the observations with simulations
of Paschos \& Norman (2005), which has a sharp reionization at $z \sim 7$.
The dashed and dotted lines are for spectral resolution of 5300 and 30000,
respectively, in the simulated spectra. The observed spectra are binned to $R\sim 2600$.
The observations appear to have reached the overlap stage at $z \sim 6$.
}
\end{figure}

Figure 11 shows the distribution of \lya dark gap lengths in the 12 lines of sight
that have Keck/ESI spectra. 
We define dark gaps as continuous regions in the quasar spectra in which {\em all}
pixels have optical depth larger than a threshold $\tau_{\rm min}$.
In the upper panel of Figure 11, we use $\tau_{\rm min} = 2.5$, 
the same as that of Songaila \& Cowie (2002) and Paschos \& Norman (2005). 
Both the average gap length and its dispersion increase dramatically at
$z>5.7$.
At this optical depth threshold,
dark regions with $\Delta r \gtrsim 40$ comoving Mpc 
already begin to appear at $z_{abs} \sim 5.3$
(e.g. Djorgovski et al. 2001).
 Long gaps
with $\Delta r \gtrsim 80$ comoving Mpc appear
at $z_{abs} \sim 5.7$; however, they are relatively rare
while most lines of sight still have
plenty of pixels with $\tau < 2.5$ and only short dark gaps.
At $z_{abs} > 5.9$, all lines of sight have long dark gaps with $\tau
> 2.5$. In many cases, the 
gap length is only a lower limit, as the gap extends all the way
to the quasar proximity zone.
Note that at the high redshift end, imperfect sky subtraction 
in regions with strong OH lines results in large residuals in the
spectrum and an {\em underestimate} of the dark gap length.
This transition from
isolated, short gaps occupying only a small fraction of pixels,
 to deep, long troughs is
the most abrupt transition among all the IGM properties we have presented in this paper,
and is difficult to explain by a gradual thickening of
the Ly$\alpha$ forest.
Rather, it suggests a phase transition in the
IGM geometry.
Detailed comparison with state-of-art simulations
will constrain the IGM geometry at this critical
stage of reionization.

Figure 11  compares the observations with the simulations
of Paschos \& Norman (2005) described in \S4.3.
The dashed and dotted lines are model predictions of the average gap length
for spectral resolution of $R=5300$ and $R=36000$. 
Our spectra are binned to $R=2600$, and should be compared to their low-resolution results. 
The observations and simulations agree very well at $z<5.5$. 
Both also show a strong break in the evolution of dark gap length;
however, the break in the observations is both  earlier (by $\Delta z \sim 0.5$)
and sharper. 
This is consistent with the comparison in Figure 9, and suggests
that the reionization epoch is at somewhat lower redshift than
in the simulations.
Note, however, the simulation of  Paschos \& Norman (2005) has a box size
of only 6.8 h$^{-1}$ comoving Mpc. They used multiple passes through the box to measure
the gap sizes larger than the box size, which could bias the result.

At $\tau_{\rm min} =2.5$ , all gaps at $z>6.0$ are longer than 100 comoving Mpc,
and most are only lower limits.
In order to further constrain the neutral fraction, we use 
a higher optical depth threshold $\tau > 3.5$ (shown in the lower panel of Figure 11).
%
Even at $z>6.0$, the average length of dark gaps is still finite:
$\Delta r \gtrsim 80$ comoving Mpc.
At this higher threshold, it is clear that the IGM is not 
yet neutral at $z\sim 6$.

In the redshift range considered in this paper, the measurement
of the dark gap distribution and the GP optical depth place similar constraints on the  
ionization of the IGM. However, at $z_{\rm abs} >6.5$, 
it will become increasingly difficult to map the reionization history
using the GP optical depth.
Currently, with $\sim 10$ hours of observing time on 10-meter class telescopes,
we are able to place lower limits on the GP optical depths 
of $\tau \lesssim 7 $ using Ly$\alpha$, and
a factor of $\sim 2$ better using \lyb and Ly$\gamma$,
corresponding to a neutral fraction of $10^{-2} - 10^{-3}$.
It is difficult to improve this measurement further in practice,
because the optical depth limit depends logarithmically on the observing time
and the ability to control sky subtraction systematics.
At $z>6$, most of the pixels are completely dark, and contain
little information. 
After entering the redshift range where continuous
dark gaps dominate, all useful information is contained in the flux
and distribution of few remaining transmitting pixels. 
Therefore, the dark gap distribution provides a powerful
alternative to map the IGM ionization at high neutral fraction.
Paschos \& Norman (2005) show that even when 
the average neutral fraction is of the order few tenths, the average gap length
is still finite, due to the presence of regions that had already been
ionized by star forming galaxies.
In the next subsection, we place an upper limit on the neutral fraction using
the finite length of the gaps.

\subsection{Constraining the Neutral Fraction with Gap Statistics}

So far in this paper, we have only considered a photoionized IGM without explicitly
taking into account the presence of strong UV ionizing sources that
caused reionization: early star forming galaxies and AGNs.
These objects create expanding HII regions even 
in a largely neutral IGM, and result in transmitting spikes in the absorption spectrum.
However, in a neutral IGM, the optical depth at the \lya resonance line
center is so large that it generates an extended GP damping wing
(\cite{M98}). Small HII regions are  
still optically thick  to the observer due to absorption from this damping wing;
and only large HII regions around long-lived, luminous sources can survive the damping
wing absorption. Therefore, the existence of transmitting regions
between dark gaps,
together with a model for the ionizing sources, can be used to place an upper limit on the IGM
neutral fraction independent of other IGM optical depth constraints.
This is similar to the constraint using the evolution of \lya emitting
galaxies (Malhotra 2005, \cite{stern05}, Haiman \& Cen 2005, Santos 2005).
In the case of \lya galaxies, a neutral IGM will significantly attenuate 
their \lya flux. Therefore, the lack of evolution in the \lya galaxy luminosity
function is used to put an upper limit on the IGM neutral fraction $\lesssim 50$\% at $z\sim 6.5$.

Haiman (2002) gives the radius of an HII region $R_s$ in a uniform IGM with
neutral fraction $f_{\rm HI}$ around a star forming galaxy with
star formation rate of $\dot{M_*}$,
\begin{equation}
R_s = 0.65 \left(\frac{\dot{M_*}}{40\ \rm M_{\odot} yr^{-1}}\right)^{1/3}
\left( \frac{t_*}{10^7 \rm yr}\right)^{1/3} f_e^{1/3} f_{\rm HI}^{-1/3}
\left( \frac{1+z}{7}\right)^{-1} \rm proper\ Mpc,
\end{equation}
where $t_*$ is the duration of the starburst,
and $f_e$ is the escape rate of Lyman Limit photons.

\begin{figure}
\epsscale{1.00}
\plotone{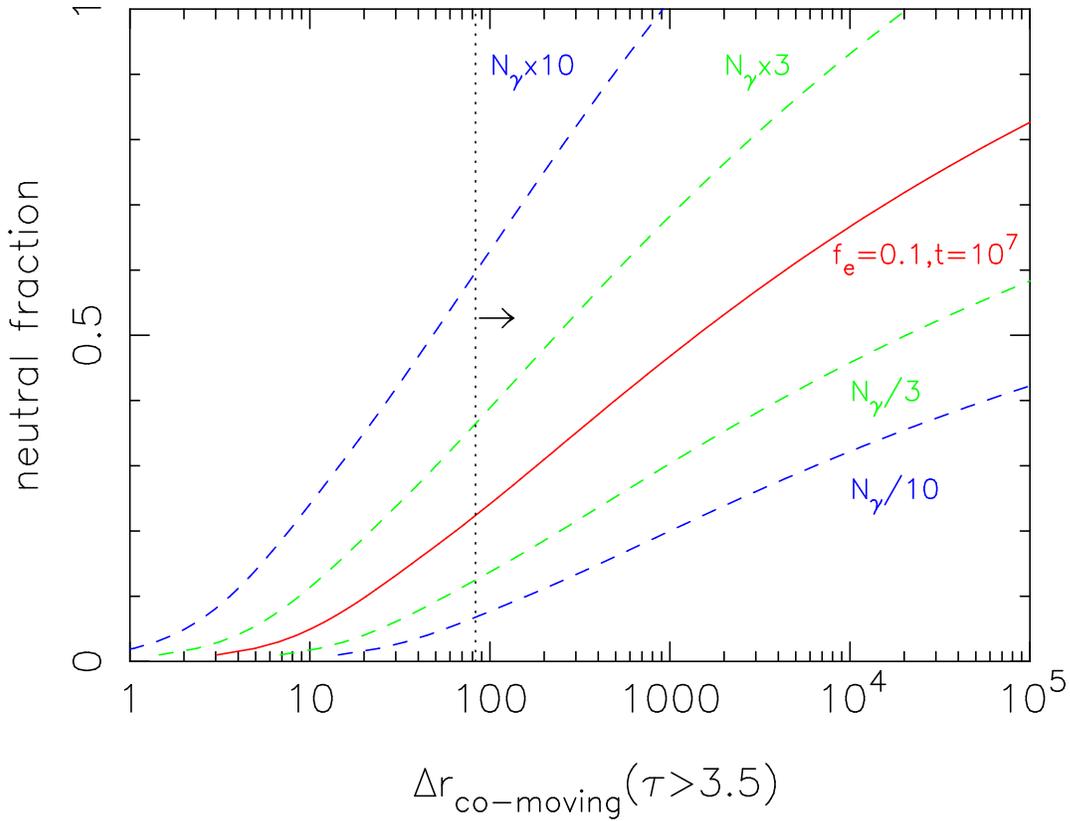}
\caption{Average length of dark gaps with $\tau > 3.5$ as a function
of IGM neutral fraction, assuming that  transmitting regions in
the quasar spectrum  are generated by HII regions surrounding 
early star forming galaxies which follow the Yan \& Windhorst (2004b)
luminosity 
function. The solid (red) curve is the default
case where the duration of star formation in each galaxy
is 10$^7$ yr and the escape fraction of ionizing photons is 10\%.
The two groups of dashed curves above and below the default case
correspond to the number of escaping photons 
3 and 10 times larger or smaller than the default case.
The dashed line is the observed limit on the average dark gap length
at $z \sim 6$, consistent with a neutral fraction $\lesssim 0.5$.}
\end{figure}

For such an HII region, the GP damping wing optical depth at
a distance $r$ from the center of the galaxy is approximately (Cen 2003c):
\begin{equation}
\tau_d(r) = 1.2 f_{\rm HI} \left(\frac{R_s-r}{1\ \rm Mpc}\right)^{-1}
\end{equation}
at large distances from the center of the resonant \lya line. 
In order to preserve a dark gap threshold $\tau_{\rm min} > 3.5$,
we require the center of the galaxy HII region to have an optical depth
from the damping wing, $\tau > 3.5$.
For a neutral IGM $f_{HI} = 1$,
we have $R_s = 0.34$ Mpc, corresponding to a star forming rate
of $\rm \dot{M_*} \sim 70\ M_{\odot} yr^{-1}$ using Eq.~(15), assuming
that the duration of star formation is $t_* = 10^7$ yr and the  escape fraction 
is  $f_e = 0.1$ (Steidel et al. 2001). This is a luminous galaxy at $z\sim 6$:
using the calibration of Bunker et al. (2004), it has a $z_{AB}$
magnitude of 24.3. Such galaxies are extremely rare at $z\sim 6$ and
are not likely to generate transmitting spikes every $\sim 100$ comoving Mpc.
The fact that the gap length is $\sim 100$ Mpc therefore already rules out a completely neutral IGM.
To include the contribution of faint galaxies and place a more  stringent constraint
on the neutral fraction, 
we assume a luminosity function of star forming galaxies at $z\sim 6$
following Yan \& Windhorst (2004b):
\begin{equation}
\Phi(M_{1300}) = 0.921 \Phi_* 10^{-0.4(\alpha+1)(M^*_{1300}-M_{1300})} \exp[-10^{0.4(M^*_{1300}-M_{1300})}],
\end{equation}
where $\Phi_* = 4.55 \times 10^{-4} \rm Mpc^{-3}$,
$\alpha=-1.6$ is the faint end slope of the luminosity function,
$M_{1300}$ is the absolute AB magnitude at rest-frame wavelength 1300\AA,
and the characteristic luminosity $M^*_{1300} = -21.03$.
The average length of the dark gap is then the distance between 
two star forming galaxies that have $\tau < 3.5$ at the center of
their respective HII regions:
\begin{equation}
\langle 1/\Delta r \rangle = \int \Phi(M_{1300}) \pi r^2_{3.5}(M_{1300}) dM_{1300},
\end{equation}
where $r_{3.5}(M_{1300})$ is the radius up to which the damping wing optical
depth is still smaller than 3.5, and the integral is over the 
luminosity range in which the \lya transition from the galaxy can survive
the damping wing absorption. 
We use the calibration of Bunker et al. (2004) to convert UV luminosity 
to star formation rate.

Figure 12 shows the gap length $\Delta r (\tau >3.5)$ as a function
of neutral fraction. In the case of $t_* = 10^7$ yr and $f_e = 0.1$,
the average gap length of $\sim 80$ comoving Mpc at $z\sim 6$ corresponds
to a neutral fraction of $\sim 0.2$. 
For a population of star forming galaxies with 10 times more photon production
($\propto f_e t_*$), the same gap length corresponds to $f_{HI} \sim 0.6$,
while for a population with 10 times less photons,
the constraint on $f_{HI} \sim 0.05$. 
The constraint in this simple model, however, is only an upper limit on
$f_{HI}$ for several reasons:
(1) we do not consider the fact that  the recombination timescale in
the early Universe is long compared to the Hubble time, and assume that all IGM ionization
is produced by HII regions around star forming galaxies. In reality,
the IGM is probably already largely ionized by previous generations of
star forming galaxies; there are highly ionized regions without active
star forming galaxies in them at the observed epoch (Wyithe \& Loeb 2005);
(2) we do not consider the clumpiness of the IGM. Early galaxies are 
formed in the highest density, most biased peaks in the density field,
and have the highest IGM density in the surrounding region. So the constraint
here is closer to the neutral fraction in the densest part of the universe,
which is higher than the average neutral fraction;
(3) we do not consider the clustering of star forming galaxies at high redshift.
If they are strongly clustered, the distance between separate clusters
of star forming galaxies is longer for the same IGM neutral fraction.
A lower IGM neutral fraction is needed to explain the observed gap length.
Therefore, we regard the constraint given in Figure 12 to be an upper limit:
the finite distance between dark gaps at $z\sim 6$ suggests that 
the IGM is not likely to be completely neutral in the regions between
HII regions ionized by early star forming galaxies.
However, a neutral fraction as high as 10--50\% is still permitted in
the optimal case.

\section{HII Regions Around Quasars}

UV photons from a luminous quasar will ionize a HII region 
(Madau \& Rees 2000, Cen \& Haiman 2000) in the IGM.
The evolution of a cosmological HII region expanding into a
neutral IGM is described by the equation (e.g., Shapiro \& Giroux 1982,
Madau, Haardt \& Rees 1999):
\begin{equation}
n_H\left(\frac{dV_I}{dt} - 3HV_I\right) = \dot{N}_Q - \frac{n_H V_I}{t_{\rm rec}},
\end{equation}
where $V_I$ is the HII region volume, and $\dot{N}_Q$ is the number of
ionizing photons the central quasar emits per unit time.
When the quasar lifetime is much less than Hubble time, as expected here,
the Hubble expansion term is not important.
When the HII region is expanding into a partially ionized IGM, as in
the case discussed in this paper, (1) the left hand side of Eq. (19) becomes
$d(f_{\rm HI}n_H V_I)/dt$, 
(2) the first term of the right hand side of Eq. (19) becomes
$\dot{N}_Q + \epsilon_B V_I$, where $\epsilon_B$ is the emissivity of
the diffuse UV background.
Assuming photoionization-recombination 
equilibrium in
the IGM outside quasar HII region: $\epsilon_B  = n_{\rm HII} n_{e} \alpha(T) =  (1-f_{\rm HI})^2 f_H / t_{\rm rec}$  (for simplicity, we assume a pure hydrogen IGM here), Eq. (19) then changes to:
\begin{equation}
\frac{d(f_{\rm HI}n_H V_I)}{dt} = \dot{N}_Q - \frac{(2-f_{\rm HI})f_{\rm HI}n_H V_I}{t_{\rm rec}}.
\end{equation}
The quasar ionizing photons are only responsible for ionizing a
fraction $f_{\rm HI}$ of
the IGM, while
the rest is ionized by the diffuse UV background
same as the case
outside quasar HII region. The factor of $(2-f_{\rm HI})$ in Eq. (20) arises
from the $(1-f_{\rm HI})^2$ dependence in photoionization-recombination equilibrium equation.
In the case of a mostly ionized IGM, $f_{\rm HI} \ll 1$, we have
\begin{equation}
\frac{d(f_{\rm HI}n_H V_I)}{dt} = \dot{N}_Q - \frac{2f_{\rm HI}n_H V_I}{t_{\rm rec}}.
\end{equation}
Therefore, for the same quasar lifetime $t$, the HII region size
scales as
\begin{equation}
R_s \propto \left(\frac{\dot{N}_Q}{2f_{\rm HI}}\right)^{1/3}.
\end{equation}
This scaling applies for both the cases in which quasar lifetime
is longer or shorter than the recombination time.
When recombination is not important (Madau \& Rees 2000, Cen \& Haiman 2000, cf. Yu \& Lu 2005), Cen \& Haiman (2002) show that
the HII region has a radius
\begin{equation}
R_s = 8.0 f_{HI}^{-1/3} (\dot{N}_Q/6.5\times 10^{57} {\rm s^{-1})^{1/3}}
(t_Q/2\times 10^7 \rm yr)^{1/3} [(1+z_Q)/7]^{-1} \rm proper\ Mpc.
\end{equation}
This is the same as Eq.~(15),  now expressing the source luminosity 
in terms of the photon production rate. 
The SDSS $z\sim 6$
quasars would produce HII regions with $R_s$ of  the order several
Mpc if the IGM were largely neutral
(e.g., Pentericci et al. 2002, White et al. 2003). 
A number of authors (Mesinger \& Haiman 2004, Wyithe \& Loeb 2004a, Wyithe et al. 2005,
Yu \& Lu 2005) used the observed size of the HII regions around a number of SDSS $z\sim 6$
quasars to constrain the IGM neutral fraction, and
suggested that the IGM could have neutral fraction $\gtrsim 0.2$
outside the quasar HII region. 
Using this size to derive the IGM neutral fraction, however, is complicated by a number of  factors:
\begin{enumerate}
\item Uncertainties in the quasar lifetime. Current observations
suggest that quasar lifetimes lie in the range 10$^5$ - 10$^8$ years
(e.g. \cite{hopkins05}, \cite{martini01}).
Most studies have assumed a lifetime of $\sim 10^7$ yrs, similar to 
the Eddington timescale for black hole growth.  
For a given $R_s$, $f_{\rm HI} \sim t_Q$ when recombination is not important. 
\item Uncertainties in the ionizing photon production rate by the quasar for a given
UV luminosity.
\item  Radiative transfer and clumpiness of
the IGM within the quasar HII region, which  will affect the 
{\em apparent} line of sight size of the HII region.
\item Contribution of ionizing photons from galaxies in the overdense
environment of luminous quasars.
\item Large scale clustering of the IGM around luminous quasars
at $z\sim 6$, which are highly biased, will affect the estimates of the IGM neutral
fraction.
\end{enumerate}

The last two factors could result in significant overestimation of
the neutral fraction. 
The SDSS quasars considered here are very rare objects
(spatial density $\sim 10^{-9}$ Mpc$^{-3}$) and thus are likely to live in
highly biased environment in the early Universe.
Fan et al. (2001) estimated that these quasars correspond to 5-6$\sigma$
peaks in the density field.
Yu \& Lu (2005) argued that up to $\sim 90$\% of the neutral hydrogen could
have already been  ionized by the star forming galaxies associated
with this high density peak.
Mesinger \& Haiman (2004) showed that by comparing high S/N, high
resolution spectra with detailed hydrodynamic simulations of quasar
HII regions, it is possible to reliably estimate the neutral fraction and
minimize the degeneracies with quasar lifetime and bolometric luminosity
along some lines of sight.

For these reasons,
the {\em absolute measurement} of the neutral fraction based on
HII region size could be off by a large factor.
Over the narrow redshift range considered here, these
systematics are likely to be roughly the same on average for all the quasars.
If there is an order of magnitude evolution in the IGM ionization,
the size of the HII region should show strong evolution,
providing a reliable {\em relative} measurement of neutral fraction.
In this section,  we study the distribution and evolution of HII region sizes 
around the nineteen $z\sim 6$ quasars
in our sample.

We first examine the definition and measurement of the HII region size
around quasars. The physical picture is that quasars are located
within an IGM with neutral fraction $f_{\rm HI}$. As the quasar 
evolves, its ionizing flux produces a surrounding region 
within which the IGM is completely ionized. 
We have shown above that for a partially ionized IGM,
the size of this proximity zone grows
as $\sim \dot{N}_Q^{1/3} t_Q^{1/3}$. 
In the case that the recombination timescale is long,
this is {\em not} a picture
of a static \strom\ sphere. When the general 
IGM is already highly ionized (valid at least
up to $z\sim 6$, as we have seen), it is not even a standard HII region.
It is also very different from the classic
foreground proximity effect at lower redshift 
(e.g. \cite{bajtlik88}, Scott et al. 2002). 
At $z<5$, the IGM ionizing background 
is much higher. At a distance $R_s$ from the quasar, the
UV background itself is much larger than the flux intensity
produced by the quasar in question.
Furthermore, the mean free path of ionizing photons is much
larger than $R_s$ calculated above. In this case, the
expanding proximity zone {\em does not} have a clear boundary.
The proximity effect is simply an excess of ionizing background
and increased ionization close to the quasar.

At $z\sim 6$, with a low UV background, the quasar
flux does produce a highly ionized proximity zone with a well-defined boundary
expanding into the general
IGM. However, the boundary of this proximity zone can be difficult 
to measure observationally. It cannot be simply 
defined as the point at which the transmitted flux reaches zero for the following reasons:
(a) if the IGM is highly ionized, not only does the transmission never
reach zero on average, it has also large fluctuations 
(\S3). Simply defining the size of the proximity zone as the point at which
the flux reaches zero will overestimate its size at low redshift.
(b) The IGM transmission is characterized by a mixture of
dark gaps and sharp transmission peaks (\S5). Using zero flux
as the boundary is affected by the resolution and smoothing length
used. (c) In the case of a neutral IGM, reaching
zero flux does not necessarily mean reaching the boundary
of the proximity zone: the IGM could still be ionized by the quasar
to a level much higher than the surrounding material, but with optical
depth higher than the detection limit ($f_{HI} > 10^{-3}$). 

\begin{figure}
\epsscale{1.00}
\plotone{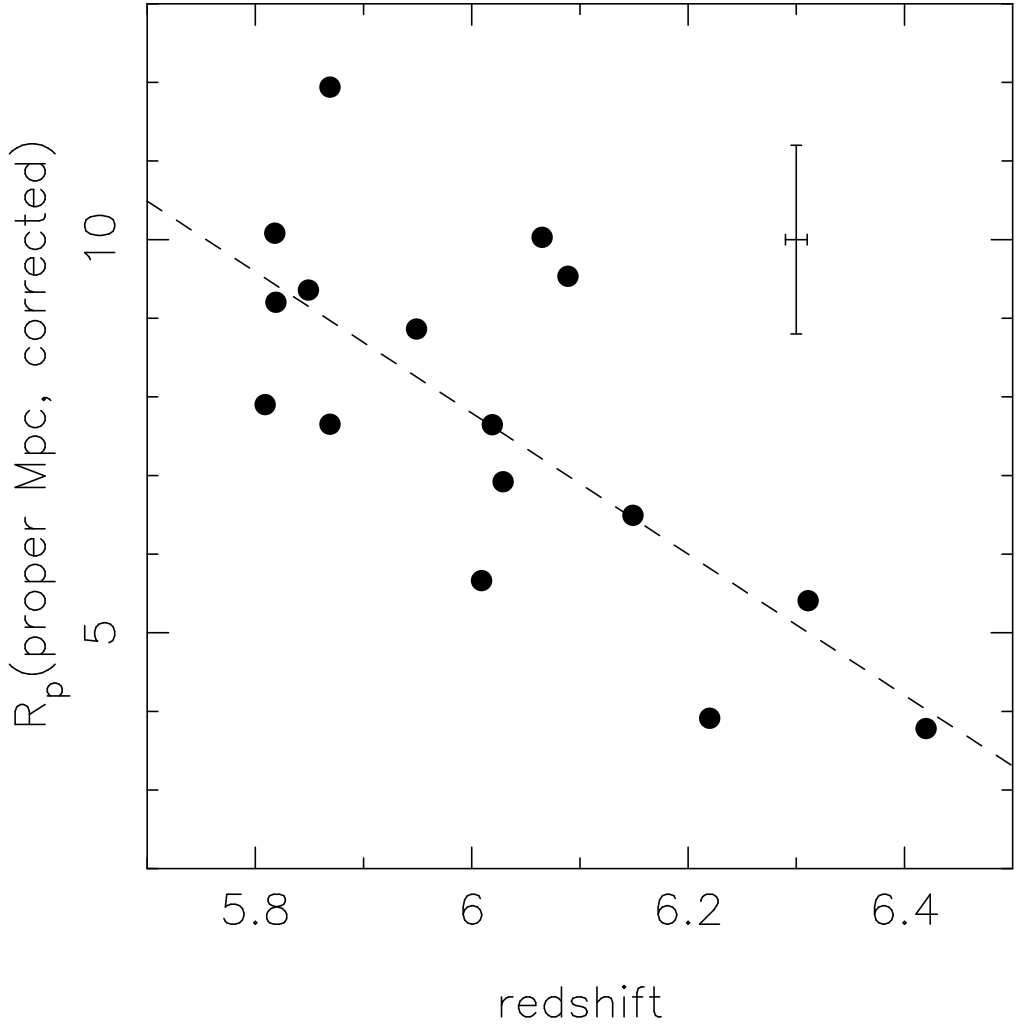}
\caption{Size of quasar proximity zone as a function of the redshift
of the quasar. The size $R_p$ is the line of sight distance from the
quasar to the point at which the transmitted flux ratio falls to 0.1 of the continuum level.
The radius also has been scaled to a common absolute magnitude ($M_{1450} = -27$).
The linear fit shows a factor of 2.8 decrease in the HII region size
around quasars from $z=5.7$ to 6.4.
The error on $R_p$ is dominated by the uncertainty in the quasar redshift. }
\end{figure}

Therefore, instead of using the \strom\ sphere radius $R_s$, we 
define the size of
the proximity zone $R_p$ as the region where the transmitted flux ratio
is above 0.1, when smoothed to a resolution of 20\AA.
For the quasars in our sample ($z>5.7$), the average IGM transmission
is smaller than 0.04. Therefore the proximity zone defined this way 
is not affected by the IGM properties {\em outside} the quasar HII region,
even with the large observed fluctuation in the IGM transmission.
To first order, at $R_p$, the IGM ionization state is the same for all the quasars,
so we have: $R_p \propto (\dot{N}_Q t_Q)^{1/3} (f_{\rm HI})^{-1/3}$.
The quasars in our sample have very similar intrinsic properties.
So even though we cannot calibrate this relation on an absolute scale,
we can use the evolution of $R_p$ to measure the {\em relative} change in
the IGM neutral fraction into which the quasar HII region is expanding.

The other uncertainty comes from the difficulty in determining the quasar redshift.
It is well known that there is a systematic velocity offset between  
high-ionization emission lines such as CIV and SiIV and the systematic redshift
of the quasar host galaxy, as measured by narrow lines such as [OIII],
or CO molecular lines in the host galaxy. 
On the other hand, low-ionization lines such as MgII show little
offset from the systematic redshift. 
Richards et al. (2002b) used a large sample of SDSS quasars to show
that CIV is blueshifted from MgII by $824\pm 511$ km s$^{-1}$.
At $z\sim 6$, we have $z_{\rm MgII} = z_{\rm CIV} + (0.019 \pm 0.012)$.
Most of the redshifts in Table 1 are determined using high-ionization
lines such as CIV, SiIV or NV. When calculating the size of the proximity zone,
this shift becomes a major source of systematic error.
Five of the quasars in our sample have MgII or CO redshifts:
J1148+5251: $z_{\rm CO} = 6.42$ (Walter et al. 2003),
J1623+3112: $z_{\rm MgII} = 6.22$ (L. Jiang et al. in preparation),
J1630+4012: $z_{\rm MgII} = 6.07$, J1030+0524: $z_{\rm MgII} = 6.31$ (Iwamuro et al. 2004), and J0836+0054: $z_{\rm MgII} = 5.82$ (Pentericci et al. 2005). 
We adopt the MgII or CO redshift for these five objects when calculating
the size of the proximity zone.
For the rest of the sample, we assume $z_{\rm MgII} = z_{\rm high-ionization} + 0.02$
in the calculation, and an uncertainty in the redshift determination of 0.02.
 
Figure 13 shows $R_p$ as a function of redshift. We have scaled
$R_p$ to a common absolute magnitude $M_{1450} = -27$ (assumed to be
proportional to $\dot{N_q}$) as
$R_p$ (corrected) $= R_p \times 10^{-0.4(-27+M_{1450})/3}$.
Correcting for luminosity differences significantly tightens the scatter
around a linear regression.
We also exclude the two BAL quasars as well as SDSS J1335+3533 ($z=5.94$)
from this test. 
The latter object is a lineless quasar presented in Fan et al. (2005),
in which we used the size of the proximity zone to estimate its redshift.
The error bars on $R_p$ are dominated by the uncertainties in the 
emission line redshift of the quasars $\sigma(z_{em}) \sim 0.02$,
yielding an uncertainty of $\sim 1.2$ Mpc.
The solid line in Figure 13 is the best linear fit of $R_p$ as a function
of redshift: 
\begin{equation}
R_p = 6.7 - 8.1 \times (z-6)\ \rm Mpc. 
\end{equation}
This linear relation has a Pearson correlation coefficient of 0.72,
which is significant at the 99.8\% level for our sample of 16 quasars.
We use this linear fit only to establish the correlation between $R_p$ and redshift,
rather than making any detailed statement about the functional dependence of
$R_p$ on $z$.
From $z=5.7$ to 6.4, the average size of the proximity zone $R_p$
has decreased from 9.1 Mpc to 3.5 Mpc, a factor of 2.8.
From Figure 13, the scatter of HII region sizes at any given
redshift is of the order of 2 Mpc,
not significantly larger than the size of the error bar.
Based on Eq.~(23), this indicates a small dispersion in
intrinsic quasar properties, such as bolometric
correction, quasar lifetime and relative bias of
the quasar environment,  at a given redshift.

\begin{figure}
\epsscale{1.00}
\plotone{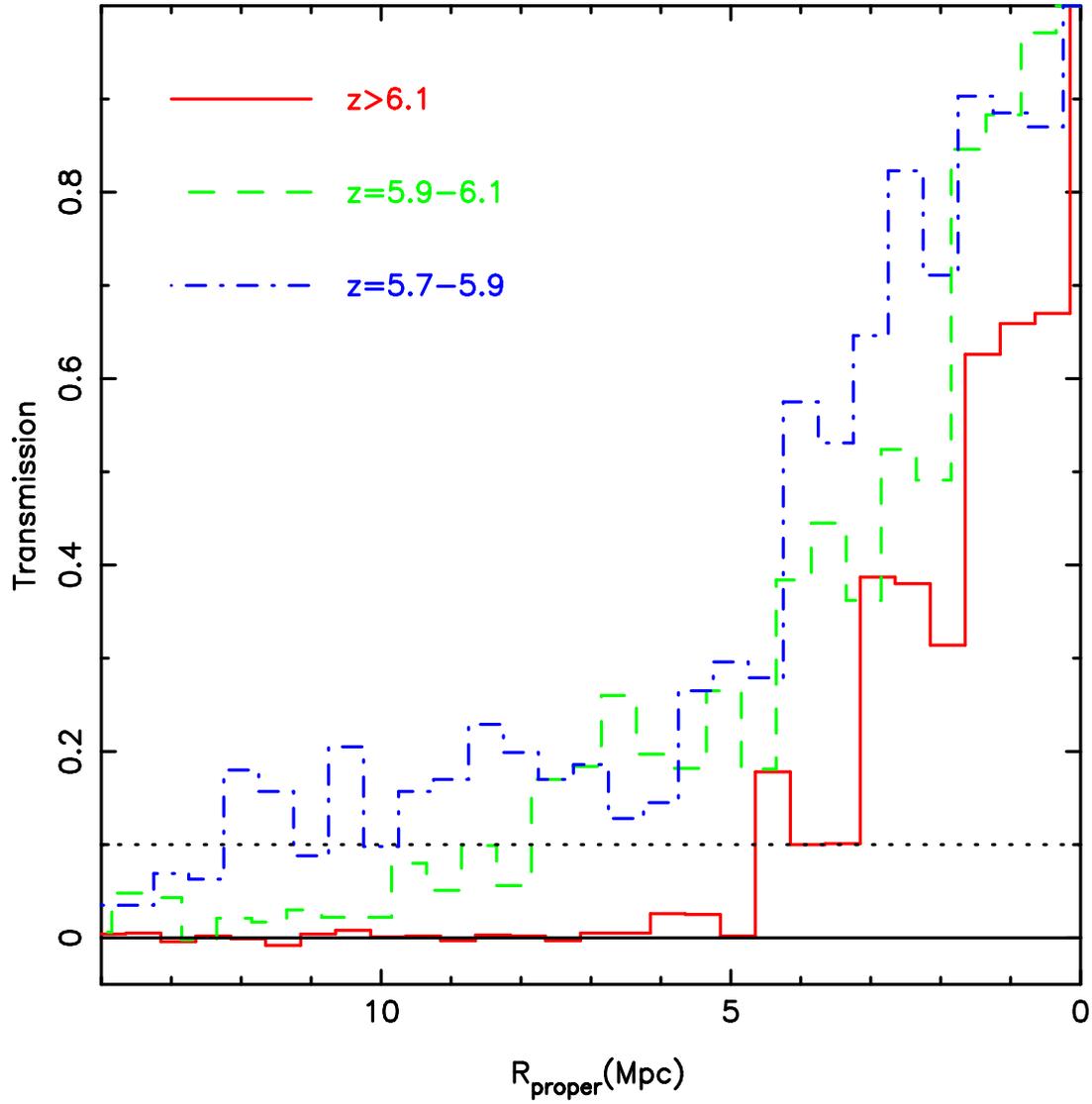}
\caption{Average absorption profiles in the proximity zone for quasars
at three different redshift bins: $z=5.7 - 5.9$, $z=5.9 - 6.1$ and
$z>6.1$. The \lya+NV emission profile has been fitted with a Gaussian and
removed. 
The absorption profile is significantly narrower at $z>6.1$ than at
lower redshifts.}
\end{figure}

Figure 14 shows the mean absorption profiles in the quasar proximity zones
as a function of redshift. We do not scale the profile according
to quasar luminosity in this case. The profile is noisy due to the
small number of quasars (4--8) in different redshift bins
and the presence of strong absorption lines within the proximity zones,
due to the clumpy IGM close to the quasar.
However, the evolutionary trend is evident: the $z=5.7 - 5.9$ profile is
the most extended; the transmission declines to $\sim 0.1$
at $\sim 8 - 10$ Mpc.
The $z=5.9 - 6.1$ profile reaches the same level at $R_p \sim 7$ Mpc,
while the profile at $z>6.1$ is the narrowest, declining 
to $\cal{T}$ = 0.1 after only $\sim 3$ Mpc.
This trend is not significantly affected by the choice of binning.
Also note that the profiles in the  low redshift bins continue to decline
beyond $R_p$ until reaching the IGM level, but the level never reaches zero flux.
Only in the highest redshift bin does the profile reach zero flux,
as the IGM outside the quasar HII region has $\tau \gtrsim 5$. 
For $z>6$, when the flux outside the HII region does reach zero,
the proximity zone size $R_p$ defined this way is $\sim$ 20\% smaller
than that using the zero-flux size.

The significant evolution of the size of quasar HII regions validates
the physical picture that quasar HII regions are expanding into
an increasingly neutral IGM at higher redshift.
Since $R_p \propto [(1+z) f_{\rm HI}]^{-1/3}$
(after scaling out the dependence on quasar luminosity), a decrease of 
HII region sizes by a factor of 2.8
from $z=5.7$ to 6.4 suggests an increase of the neutral fraction by a factor
of 14.
Note that this relation applies when $f_{\rm HI} \ll 1$, as in the case 
here; at larger nuetral fraction, a detailed solution of Eq. (20) is
required.
This constraint on the {\em relative} increase of the neutral fraction
does not depend on the model of photoionization and IGM density distribution
we used in \S3. 

Mesinger \& Haiman (2004) examined the absorption profile in the
line of sight of SDSS J1030+0534 ($z=6.28$). 
In this line of sight, the \lyb transition reaches zero flux at a redshift
$\sim 0.02$ lower than that of the \lya transition. They model
this feature as the boundary of a \strom\ sphere, and suggest that
the size of this \strom\ sphere is consistent with a largely neutral
IGM.
From Figure 10, we clearly see the extra transmission in \lyb
for SDSS J1030+0524. 
However, we do not
observe this feature in any other line of sight. 

Our GP optical depth measurements show that at $z_{abs} \sim 5.7$,
the volume- and mass-averaged neutral fractions are 
$\sim 9.3\times 10^{-5}$ and $2.8\times 10^{-3}$ respectively (see Figure 7).
Our result here implies that the neutral fraction has increased by
a factor of 14 at $z\sim 6.4$, to $1.3\times 10^{-3}$ and 0.04, respectively.
Note that the high mass averaged neutral fraction of 0.04 is lower 
than several previous estimates (Mesinger \& Haiman 2004, Wyithe et al. 2005)
based on measurements of HII region size,
but still is in general agreement. 
Our estimate of neutral fraction evolution is based on a simple model
in which a uniform UV background ionizes the IGM to a neutral fraction
of $f_{\rm HI}$, and the quasar provides the extra UV photons that
produce a completely ionized HII region. Yu \& Lu (2005) describe
a more detailed, self-consistent model that includes ionization from
both stellar sources and quasars. Their model considers in detail
the influence of the bias factor of quasar environment and quasar lifetime on the size
of the quasar HII region. Such detailed modeling is beyond the scope of
this paper.
\section{Discussion and Summary}

The goal of this paper is to study the evolution of the IGM ionization state near
the end of reionization using absorption spectra of the highest redshift quasars
known. We use three types of measurements to study IGM properties.
We now discuss their advantages and disadvantages,
with an eye towards application to yet higher redshift.

\noindent
{\bf 1. Evolution of Gunn-Peterson optical depth.}
In \S3, we measured the GP optical depth in the \lya, \lyb and \lyc transitions.
This is directly linked to the neutral
hydrogen density of the IGM. Using a photoionization model, we can compute
the underlying UV ionizing background, and estimate the neutral fraction and
mean free path of UV photons. However, as the optical depth increases to $\tau \gtrsim 5$
at $z>6$, the \lya transition becomes completely saturated. It is difficult to place stronger
constraints as the measurement begins to be dominated by systematic observational errors
in the red and near-IR part of the spectrum.  The \lyb and \lyc transitions (\S3.2) give
more stringent constraints due to their smaller oscillator strengths, but they can only
improve the limit by a factor of $2-4$. In addition, one has to correct for the foreground
absorption from lower order transitions. 
Direct measurement of the GP optical depth can only constrain
the neutral fraction of the IGM to lower limits in the range $\sim 10^{-3}$ to $10^{-2}$. 
Since at high redshift, all IGM transmission is through a small number of transmitting
spikes with high ionization and/or low density, the {\em effective} optical depth is
actually {\em not} sensitive to the region of the IGM where most of the neutral hydrogen
lies. 
Furthermore, the interpretation of optical depth depends on the photoionization model 
and IGM density distribution models used. 
The exact values of derived ionization parameters, such as ionizing background and neutral fraction,
are therefore model-dependent. 
But for a narrow redshift range, our calculation shows a robust strong evolutionary
trend on the ionization parameters.
Thus using the evolution of GP optical depth is most useful at the low redshift
end, and it begins to lose its usefulness when the neutral fraction increases to $\sim 10^{-3}$.

\noindent
{\bf 2. Distribution of Gunn-Peterson troughs and dark gaps.}
In \S4, we discussed using the distribution of 
absorption gaps to characterize the evolution of the IGM.
Qualitatively, these statistics reveal very similar trends to the GP optical depth
measurements. But they also contain high order information, related to the clustering
of transmission spikes and dark gaps. 
The evolution of dark gaps appears to be the most sensitive quantity tracing
the reionization process: of the statistics we examine, it shows the
most dramatic evolution at $z>6$. 
The distribution of dark gap length is readily computable from both observations and cosmological simulations. 
Simulations show that even when the average neutral fraction is high,
the dark gap length is still finite due to the presence of regions that were
ionized by star-forming galaxies.
In \S5.3, we used the length of dark gaps and the observed galaxy
luminosity function at $z\sim 6$ to put an independent
upper limit on the  IGM neutral fraction. 
Of course, these statistics are not straightforward to calculate from analytic
or semi-analytic models, and must  be derived from detailed simulations
that include the clustering of the IGM and ionizing sources, and
take into account spectral resolution and S/N of observed spectra.

\noindent
{\bf 3. HII regions around high-redshift sources.}
In \S6, we measured the  HII region sizes 
around high-redshift quasars, and showed that the sizes decrease significantly
with redshift, consistent with a rapid increase in the IGM neutral fraction.
This method provides a powerful and independent way of measuring the IGM neutral fraction,
and can in principle be applied to regions with high neutral fraction if
luminous sources can be discovered.
The disadvantage is that the sizes of HII regions are quite model dependent:
theoretical calculations require not only knowledge of IGM properties, but also
the intrinsic properties of quasars, including their lifetime and bias relative to the
density field at high redshift. 
Further theoretical modelling will improve this constraint. 
As these quantities are unlikely to change over the redshift range we probe,
we  used the HII region sizes as a robust measure of the {\em fractional}
change in the neutral fraction over this redshift range.

In \S1, we posed three questions regarding the evolution of the IGM and the end of
reionization. We now present our major conclusions based on our analysis
of this sample of nineteen $z>5.7$ quasars.

\begin{enumerate}
\item There is a strong evolution of the IGM ionization state at high redshift,
in terms of the effective GP optical depth, the derived UV background, the neutral fraction,
the mean free path of UV photons, and the extent of dark gaps.
This evolution  {\em accelerates}
at $z>5.7$. 
The GP optical depth evolution changes from $\tau^{\rm eff}_{\rm GP} \sim (1+z)^{4.3}$
to $(1+z)^{\gtrsim 11}$, indicating a lower limit in the  increase in the volume-averaged
neutral fraction of a factor of  7 between $z=5.5$ and 6.2.

\item Accompanying the increased average neutrality of the IGM is a rapid increase
in the relative dispersion of IGM properties along different lines of sight.
We find that this increased dispersion implies 
fluctuations by a factor of $\gtrsim 4$ in the UV background at $z>6$.
This is not surprising, as the mean free path of UV photons at $z\sim 6$ is  comparable
to the correlation length of star forming galaxies which likely provide most
of the photons that ionized the Universe.

\item The most effective measure of the IGM evolution at high redshift is the distribution
of dark absorption gaps in the IGM. The average length of dark gaps
shows a dramatic increase at $z\sim 6$. This quantity shows the largest
line of sight variations: dark gaps as long as 50 comoving Mpc appear
by $z\sim 5.5$, while some lines of sight at $z>6$ are still somewhat transparent.

\item Using the evolution of HII region sizes around the quasars,
we find the neutral fraction
of the IGM has increased by $\sim 14$ between $z=5.5$ and 6.4,
consistent with our GP optical depth estimates. This implies
a mass-averaged neutral fraction as high as $\sim 4$\% in regions around luminous quasars at $z\sim 6.4$.

\item There is no compelling evidence that the IGM at $z\sim 6.4$ is
largely neutral. On the contrary, the finite length of dark gaps in the spectra
implies an upper limit on the neutral fraction less than 50\%, likely even lower.

\end{enumerate}

\begin{figure}
\epsscale{1.00}
\plotone{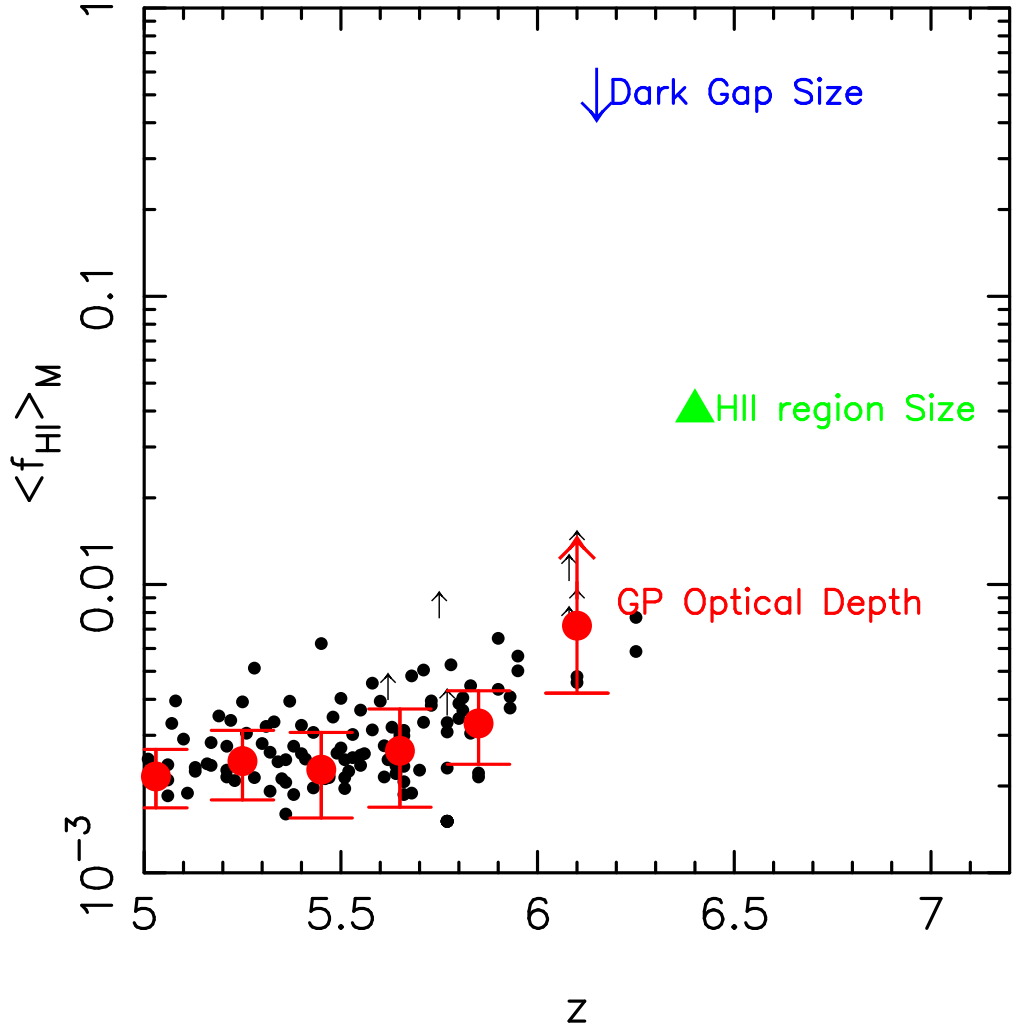}
\caption{The evolution of the mass-averaged neutral fraction using constraints
from GP optical depth measurements, HII region size evolution, and
distribution of dark absorption gaps.}
\end{figure}

Figure 15 summarizes our constraints on the 
mass-averaged neutral fraction using quasar spectra presented in this paper.
The analysis here confirms the main result of Paper I based on only four quasars:
the detections of deep GP troughs indicate an accelerated rate of IGM
evolution at $z>5.7$, consistent with the IGM transition at the end of the overlapping
stage of reionization. However, the results do not indicate that the IGM
has achieved a high level of neutrality at $z\sim 6$.
The large dispersion of the IGM properties
along different lines of sight  strongly suggests that the reionization
process is complex and not likely a uniform phase transition over a very narrow
redshift range. 

If luminous sources at $z\sim 6 - 10$ can be discovered
by JWST and next generation wide-field IR surveys, 
statistics of dark gaps in their absorption spectra and measurements of HII region
sizes surrounding these sources will provide the most useful probe of the history
of reionization using high-redshift discrete sources, and complement constraints
from CMB polarization measurements (e.g. PLANCK) and 21cm experiments
(e.g. LOFAR, MWA, PAST and SKA).

Funding for the Sloan Digital Sky Survey (SDSS) has been provided by the Alfred P. Sloan Foundation, the Participating Institutions, the National Aeronautics and Space Administration, the National Science Foundation, the U.S. Department of Energy, the Japanese Monbukagakusho, the Max Planck Society, and the HEFCE.
The SDSS is a joint project of The University of Chicago, Fermilab, the Institute for Advanced Study, the Japan Participation Group, The Johns Hopkins University, the Korean Scientist Group, Los Alamos National Laboratory, the Max-Planck-Institute for Astronomy (MPIA), the Max-Planck-Institute for Astrophysics (MPA), New Mexico State University, University of Pittsburgh, University of Portsmouth, Princeton University, the United States Naval Observatory, and the University of Washington.
We thank Andrea Ferrara, Doug Finkbeiner, Steve Furlanetto, Nick Gnedin, Zoltan Haiman, Avery Meiksin, Mike Norman, Peng Oh, 
Stuart Wyithe,  and Idit Zehavi for helpful discussions,
and the referee for valuable comments and suggestions.
XF thanks KITP for hospitality during part of this project.
We acknowledge support from NSF grant AST 03-07384, a Sloan Research Fellowship,
a Packard Fellowship for Science and Engineering (X.F.), NSF grant
AST 03-07409 (M.A.S., )
the Institute of Geophysics and Planetary Physics (operated
under the auspices of the US Department of Energy by
Lawerence Livermore National Laboratory under contract
No. W-7045-Eng-48) (R. H. B.) and NSF grant AST 03-07582 (D.P.S.).


\newpage

\begin{deluxetable}{lccccc}
\tablenum{1}
\tablecolumns{6}
\tablecaption{Spectroscopic Observations of the $z\sim 6$ Quasar Sample}
\tablehead
{
Quasar (SDSS) & redshift & $M_{1450}$ & Telescope$^1$ & Exposure & Ref$^2$ \\
& & & & (hours) & 
}
\startdata
J114816.64$+$525150.3 & 6.42 & --27.8  & Keck & 11.5 & 4,5 \\
J103027.10$+$052455.0 & 6.28 & --27.2 & Keck & 10.3 & 2,5 \\
J162331.81$+$311200.5 & 6.22 & --26.6 & Keck & 1.0  & 6,8 \\
J104845.05$+$463718.3$^3$ & 6.20 & --27.6 & Keck & 4.5  & 4,8 \\
J125051.93$+$313021.9 & 6.13 & --27.1 & Keck & 1.0  & 7,7 \\
J160253.98$+$422824.9$^4$ & 6.07 & --26.8 & MMT  & 1.3  & 6,6 \\
J163033.90$+$401209.6 & 6.05 & --26.1 & HET  & 1.0  & 4,4 \\
J113717.73$+$354956.9 & 6.01 & --27.1 & Keck & 0.7  & 7,7 \\
J081827.40$+$172251.8 & 6.00 & --27.4 & Keck & 0.3  & 7,7 \\
J130608.26$+$035626.3 & 5.99 & --26.9 & Keck & 0.2  & 2,3 \\
J133550.81$+$353315.8 & 5.95 & --26.8 & MMT  & 2.0  & 7,7 \\
J141111.29$+$121737.4 & 5.93 & --26.8 & Keck & 1.0  & 6,8 \\
J084035.09$+$562419.9 & 5.85 & --26.6 & MMT  & 1.0  & 7,7 \\
J000552.34$-$000655.8 & 5.85 & --26.5 & Keck & 0.3  & 6,8 \\
J143611.74$+$500706.9 & 5.83 & --26.3 & MMT  & 1.0  & 7,7 \\
J083643.85$+$005453.3 & 5.82 & --27.8 & Keck & 0.3  & 2,3 \\
J000239.39$+$255034.8& 5.80 & --27.7 & MMT  & 1.0  & 6,6 \\
J092721.82$+$200123.7 & 5.79 & --26.8 & KP   & 3.7  & 7,7 \\
J104433.04$-$012502.2$^3$& 5.74 & --27.5 & Keck & 0.7  & 1,1
\enddata
\tablenotetext{}{1. Telescopes and instrument: Keck: Keck/ESI;
MMT: MMT/Red Channel; KP: Kitt Peak 4-meter/MARS}
\tablenotetext{}{ 2. References of discovery spectrum and spectrum
presented here; (1). Fan et al. 2000,
(2). Fan et al. 2001,
(3)  Becker et al. 2001,
(4). Fan et al. 2003,
(5). White et al. 2003,
(6). Fan et al. 2004,
(7). Fan et al. 2005,
(8). This paper}
\tablenotetext{}{3. BAL quasar}
\tablenotetext{}{4. This object was listed as J160254.18+422822.9 in
Fan et al. (2004); the position given here reflects improved astrometry.}
\end{deluxetable}

\newpage
\begin{deluxetable}{cccr}
\tablewidth{0pt}
\tablenum{2}
\tablecolumns{4}
\tablecaption{\lya Transmitted Flux Ratio}
\tablehead
{
Quasar & $z_{em}$ & $ z_{abs}$ & $\cal{T}$ \hspace{1cm} 
}
\startdata
J0002$+$2550 & 5.80 & 5.58 &   0.0170 $\pm$   0.0062 \\ 
 & & 5.43 &   0.0573 $\pm$   0.0066 \\ 
 & & 5.28 &   0.0205 $\pm$   0.0045 \\ 
 & & 5.13 &   0.1243 $\pm$   0.0050 \\ 
 & & 4.98 &   0.1002 $\pm$   0.0050 \\ \hline 
J0005$-$0006 & 5.85 & 5.64 &   0.0823 $\pm$   0.0069 \\ 
 & & 5.49 &   0.0718 $\pm$   0.0070 \\ 
 & & 5.34 &   0.0961 $\pm$   0.0066 \\ 
 & & 5.19 &   0.0578 $\pm$   0.0063 \\ 
 & & 5.04 &   0.1567 $\pm$   0.0073 \\ \hline 
J0818$+$1722 & 6.00 & 5.81 &   0.0216 $\pm$   0.0053 \\ 
 & & 5.66 &   0.0440 $\pm$   0.0040 \\ 
 & & 5.51 &   0.0984 $\pm$   0.0043 \\ 
 & & 5.36 &   0.1192 $\pm$   0.0039 \\ 
 & & 5.21 &   0.0884 $\pm$   0.0039 \\ 
 & & 5.06 &   0.1285 $\pm$   0.0042 \\ \hline 
J0836$+$0054 & 5.82 & 5.52 &   0.0907 $\pm$   0.0011 \\ 
 & & 5.37 &   0.0348 $\pm$   0.0009 \\ 
 & & 5.22 &   0.0606 $\pm$   0.0009 \\ 
 & & 5.07 &   0.0751 $\pm$   0.0011 \\ 
 & & 4.92 &   0.1276 $\pm$   0.0011 \\ \hline 
J0840$+$5624 & 5.85 & 5.66 &   0.0883 $\pm$   0.0176 \\ 
 & & 5.51 &   0.1127 $\pm$   0.0260 \\ 
 & & 5.36 &   0.1661 $\pm$   0.0202 \\ 
 & & 5.21 &   0.1191 $\pm$   0.0167 \\ 
 & & 5.06 &   0.1765 $\pm$   0.0190 \\ \hline 
J0927$+$2001 & 5.79 & 5.61 &   0.0884 $\pm$   0.0136 \\ 
 & & 5.46 &   0.1041 $\pm$   0.0163 \\ 
 & & 5.31 &   0.0596 $\pm$   0.0104 \\ 
 & & 5.16 &   0.1165 $\pm$   0.0105 \\ 
 & & 5.01 &   0.1268 $\pm$   0.0117 \\ \hline 
J1030$+$0524 & 6.28 & 6.10 &   0.0012 $\pm$   0.0010 \\ 
 & & 5.95 &   0.0060 $\pm$   0.0010 \\ 
 & & 5.80 &   0.0260 $\pm$   0.0012 \\ 
 & & 5.65 &   0.0462 $\pm$   0.0009 \\ 
 & & 5.50 &   0.0661 $\pm$   0.0009 \\ 
 & & 5.35 &   0.1147 $\pm$   0.0008 \\ 
\enddata
\end{deluxetable}

\newpage
\begin{deluxetable}{cccr}
\tablewidth{0pt}
\tablenum{2}
\tablecolumns{4}
\tablecaption{\lya Transmitted Flux Ratio (continued)}
\tablehead
{
Quasar & $z_{em}$ & $z_{abs}$ & $\cal{T}$ \hspace{1cm}
}
\startdata
J1044$-$0125 & 5.74 & 5.55 &   0.0686 $\pm$   0.0022 \\ 
 & & 5.40 &   0.0520 $\pm$   0.0020 \\ 
 & & 5.25 &   0.0427 $\pm$   0.0019 \\ 
 & & 5.10 &   0.0898 $\pm$   0.0020 \\ 
 & & 4.95 &   0.1139 $\pm$   0.0022 \\ \hline 
J1048$+$4637 & 6.20 & 5.68 &   0.0117 $\pm$   0.0011 \\ 
 & & 5.53 &   0.0519 $\pm$   0.0012 \\ 
 & & 5.38 &   0.0736 $\pm$   0.0011 \\ \hline 
J1137$+$3549 & 6.01 & 5.83 &   0.0116 $\pm$   0.0029 \\ 
 & & 5.68 &   0.1010 $\pm$   0.0024 \\ 
 & & 5.53 &   0.0742 $\pm$   0.0026 \\ 
 & & 5.38 &   0.1341 $\pm$   0.0022 \\ 
 & & 5.23 &   0.1323 $\pm$   0.0023 \\ 
 & & 5.08 &   0.0530 $\pm$   0.0025 \\ \hline 
J1148$+$5251 & 6.42 & 6.25 &   0.0015 $\pm$   0.0005 \\ 
 & & 6.10 &   0.0051 $\pm$   0.0005 \\ 
 & & 5.95 &   0.0038 $\pm$   0.0005 \\ 
 & & 5.80 &   0.0186 $\pm$   0.0006 \\ 
 & & 5.65 &   0.0433 $\pm$   0.0005 \\ 
 & & 5.50 &   0.0278 $\pm$   0.0005 \\ \hline 
J1250$+$3130 & 6.13 & 5.90 &   0.0108 $\pm$   0.0033 \\ 
 & & 5.75 &   0.0055 $\pm$   0.0030 \\ 
 & & 5.60 &   0.0248 $\pm$   0.0026 \\ 
 & & 5.45 &   0.0077 $\pm$   0.0026 \\ 
 & & 5.30 &   0.0776 $\pm$   0.0024 \\ \hline 
J1306$+$0356 & 5.99 & 5.77 &   0.0645 $\pm$   0.0020 \\ 
 & & 5.62 &   0.0690 $\pm$   0.0016 \\ 
 & & 5.47 &   0.0991 $\pm$   0.0015 \\ 
 & & 5.32 &   0.0864 $\pm$   0.0014 \\ 
 & & 5.17 &   0.1156 $\pm$   0.0013 \\ 
\enddata
\end{deluxetable}

\newpage
\begin{deluxetable}{cccr}
\tablewidth{0pt}
\tablenum{2}
\tablecolumns{4}
\tablecaption{\lya Transmitted Flux Ratio (continued)}
\tablehead
{
Quasar & $z_{em}$ & $z_{abs}$ & $\cal{T}$ \hspace{1cm}
}
\startdata

J1335$+$3533 & 5.94 & 5.73 &   0.0224 $\pm$   0.0059 \\ 
 & & 5.58 &   0.0445 $\pm$   0.0074 \\ 
 & & 5.43 &   0.1215 $\pm$   0.0083 \\ 
 & & 5.28 &   0.1217 $\pm$   0.0058 \\ 
 & & 5.13 &   0.1293 $\pm$   0.0064 \\ \hline 
J1411$+$1217 & 5.93 & 5.71 &   0.0322 $\pm$   0.0033 \\ 
 & & 5.56 &   0.0665 $\pm$   0.0031 \\ 
 & & 5.41 &   0.0858 $\pm$   0.0029 \\ 
 & & 5.26 &   0.0690 $\pm$   0.0028 \\ 
 & & 5.11 &   0.1650 $\pm$   0.0030 \\ \hline 
J1436$+$5007 & 5.83 & 5.66 &   0.0714 $\pm$   0.0217 \\ 
 & & 5.51 &   0.0775 $\pm$   0.0315 \\ 
 & & 5.36 &   0.0895 $\pm$   0.0239 \\ 
 & & 5.21 &   0.1292 $\pm$   0.0199 \\ 
 & & 5.06 &   0.1509 $\pm$   0.0224 \\ \hline 
J1602$+$4228 & 6.07 & 5.85 &   0.0687 $\pm$   0.0057 \\ 
 & & 5.70 &   0.0729 $\pm$   0.0044 \\ 
 & & 5.55 &   0.0795 $\pm$   0.0058 \\ 
 & & 5.40 &   0.0802 $\pm$   0.0055 \\ 
 & & 5.25 &   0.0934 $\pm$   0.0036 \\ \hline 
J1623$+$3112 & 6.22 & 6.08 &  --0.0071 $\pm$   0.0020 \\ 
 & & 5.93 &   0.0125 $\pm$   0.0022 \\ 
 & & 5.78 &   0.0071 $\pm$   0.0024 \\ 
 & & 5.63 &   0.0402 $\pm$   0.0018 \\ 
 & & 5.48 &   0.0407 $\pm$   0.0017 \\ 
 & & 5.33 &   0.0546 $\pm$   0.0016 \\ \hline 
J1630$+$4012 & 6.05 & 5.77 &  --0.0165 $\pm$   0.0342 \\ 
 & & 5.62 &   0.0495 $\pm$   0.0323 \\ 
 & & 5.47 &   0.1015 $\pm$   0.0353 \\ 
 & & 5.32 &   0.1376 $\pm$   0.0296 \\ 
 & & 5.17 &   0.0869 $\pm$   0.0219 \\ 
\enddata
\end{deluxetable}

\newpage
\tablewidth{0pt}
\begin{deluxetable}{cccr}
\tablenum{3}
\tablecolumns{4}
\tablecaption{\lyb Transmitted Flux Ratio}
\tablehead
{
Quasar & $z_{em}$ & $z_{abs}$ & $\cal{T}$ \hspace{1cm}
}
\startdata
J0002$+$2550 &  5.80 &
5.58  &   0.0557 $\pm$   0.0039 \\ 
J0005$-$0006 &  5.85 &
5.64  &   0.0786 $\pm$   0.0066 \\ 
J0818$+$1722 &  6.00 &
5.81  &   0.0341 $\pm$   0.0035 \\ 
J0840$+$5624 &  5.85 &
5.66  &   0.0898 $\pm$   0.0127 \\ 
J0927$+$2001 &  5.79 &
5.61  &   0.0715 $\pm$   0.0088 \\ 
J1030$+$0524 &  6.28 &
6.10  &  --0.0005 $\pm$   0.0010 \\ 
J1044$-$0125 &  5.74 &
5.55  &   0.0584 $\pm$   0.0019 \\ 
J1137$+$3549 &  6.01 &
5.83  &   0.0461 $\pm$   0.0021 \\ 
J1148$+$5251 &  6.42 &
6.25  &   0.0038 $\pm$   0.0005 \\
& & 6.10  &   0.0165 $\pm$   0.0005 \\ 
J1250$+$3130 &  6.13 &
5.90  &   0.0128 $\pm$   0.0024 \\ 
J1306$+$0356 &  5.99 &
5.77  &   0.0458 $\pm$   0.0014 \\ 
J1335$+$3533 &  5.94 &
5.73  &   0.0406 $\pm$   0.0058 \\ 
J1411$+$1217 &  5.93 &
5.71  &   0.0300 $\pm$   0.0028 \\ 
J1436$+$5007 &  5.83 &
5.66  &   0.0584 $\pm$   0.0180 \\ 
J1602$+$4228 &  6.07 &
5.85  &   0.0607 $\pm$   0.0035 \\ 
J1623$+$3112 &  6.22 &
6.08  &  -0.0036 $\pm$   0.0019 \\
& & 5.93  &   0.0305 $\pm$   0.0017 \\
J1630$+$4012 &  6.05 &
5.77  &   0.0499 $\pm$   0.0194 \\ 
\enddata
\tablenotetext{}{This is the total observed transmission, without
correcting for the foreground \lya absorption.}
\end{deluxetable}

\newpage

\tablewidth{0pt}
\newpage

\newpage
\begin{deluxetable}{cccc}
\tablenum{4}
\tablecolumns{4}
\tablecaption{\lyc GP Optical Depth Measurements}
\tablehead
{
object & $z_{em}$ & redshift range & $\tau_{\rm GP}$ using \lyc
} 
\startdata
J1030+0524 & 6.28 & 6.11 -- 6.17 & $>14.2^1$ \\
J1148+5251 & 6.42 & 6.25 -- 6.32 & $14.7 \pm 1.1$\\
J1623+3112 & 6.22 & 6.05 -- 6.15 & $>12.2$
\enddata
\tablenotetext{}{1. 2-$\sigma$ lower limit}
\end{deluxetable}

\begin{deluxetable}{lcc}
\tablenum{5}
\tablecolumns{3}
\tablecaption{Evolution of Average GP Optical Depth and Dispersion}
\tablehead
{
redshift range & $\langle \tau_{\rm GP} \rangle$ & $\sigma(\tau_{\rm GP})$
}
\startdata
4.90 -- 5.15  &2.1  &0.3 \\
5.15 -- 5.35  &2.5  &0.5 \\
5.35 -- 5.55  &2.6  &0.6 \\
5.55 -- 5.75  &3.2  &0.8 \\
5.75 -- 5.95$^1$  &4.0  & 0.8 \\
5.95 -- 6.25$^1$  &7.1  &2.1
\enddata
\tablenotetext{}{1. The redshift bin includes complete GP troughs, so the
value is a lower limit}
\end{deluxetable}

\begin{deluxetable}{cccccc}
\tablenum{6}
\tablecolumns{5}
\tablecaption{Dark Troughs with $\Delta z > 0.1$ in the Highest-redshift Quasars}
\tablehead
{
object & $z_{em}$ & redshift range & $\tau_{\rm GP}$ & $\tau_{\rm GP}$ \\
       &          &                & using Ly$\alpha$  & using \lyb
}
\startdata
J1148+5251 & 6.42 & 6.10 -- 6.32 & 6.2$\pm$0.2 & 7.7$\pm$0.2 \\
J1148+5251 & 6.42 & 5.91 -- 6.07 & 6.1$\pm$0.2 & 5.5$\pm$0.1 \\
J1030+0524 & 6.28 & 5.98 -- 6.17 & $> 6.3^1$ & $> 10.2$ \\
J1623+3112 & 6.22 & 5.96 -- 6.13 & $>5.4$ & $>8.2$ \\
J1048+4637 & 6.20 & 5.63 -- 5.75 & $>5.4$ & $>5.7$ \\
J1250+3130 & 6.13 & 5.69 -- 5.85 & $>4.6$ & 6.5$\pm$0.6 
\enddata
\tablenotetext{}{1. 2-$\sigma$ lower limit}
\end{deluxetable}

 \end{document}